\documentclass{article}

\usepackage{arxiv}

\usepackage{hyperref}       
\usepackage{xcolor}
\hypersetup{colorlinks=true, linkcolor=blue, citecolor=blue}
\usepackage{amsfonts}       
\usepackage{amsmath}
\usepackage{amssymb}
\usepackage{nicefrac}       
\usepackage{microtype}      
\usepackage{lipsum}         
\usepackage{graphicx}
\usepackage{natbib}
\usepackage{doi}
\usepackage{siunitx}
\newcommand{\bm}[1]{{\mbox{\boldmath $#1$}}}

\title{Compressing fluid flows with nonlinear machine learning: mode decomposition, latent modeling, and flow control}
\renewcommand{\shorttitle}{Fukagata \& Fukami / Compressing fluid flows with nonlinear machine learning}

\author{Koji Fukagata \\
	Department of Mechanical Engineering,
	Keio University\\
	fukagata@mech.keio.ac.jp \\
	\And
 Kai Fukami \\
	Department of Aerospace Engineering,
	Tohoku University\\
	kfukami1@tohoku.ac.jp\\
}

\begin{document}
\maketitle

\begin{abstract}
An autoencoder is a self-supervised machine-learning network trained to output a quantity identical to the input.
Owing to its structure possessing a bottleneck with a lower dimension, an autoencoder works to achieve data compression, extracting the essence of the high-dimensional data into the resulting latent space.
We review the fundamentals of flow field compression using convolutional neural network-based autoencoder (CNN-AE) and its applications to various fluid dynamics problems.
We cover the structure and the working principle of CNN-AE with an example of unsteady flows while examining the theoretical similarities between linear and nonlinear compression techniques. 
Representative applications of CNN-AE to various flow problems, such as mode decomposition, latent modeling, and flow control, are discussed.
Throughout the present review, we show how the outcomes from the nonlinear machine-learning-based compression may support modeling and understanding a range of fluid mechanics problems.
\end{abstract}

\vspace{2pc}

\section{Introduction}

{
We are still in the midst of the third artificial intelligence (AI) boom, which started around 2006.
The recent development of AI looks even accelerated further with the spread of generative AI, as exemplified by Stable Diffusion, ChatGPT, and DeepSeek.
Among different AI techniques, machine learning has played a central role in the third AI boom.
Examples include facial image recognition and object detection.
Compared to the time of the second AI boom in the 1980s, it has become easier to perform research on machine learning with the support of easy-to-use libraries and hardware.
In addition, the easier access to big data may also be regarded as a factor accelerating the democratization of machine learning.
Applications of machine-learning techniques have attracted considerable attention in a range of problems in science and engineering, as symbolized by the 2024 Nobel Prizes in Physics and Chemistry.

Machine learning has also shown potential in the fluid mechanics field for supporting the understanding and modeling of spatiotemporal complex behavior (Brunton {\it et al}~2020, Vinuesa {\it et al}~2023).
For example, machine-learning-based modeling in turbulence simulations for Reynolds-averaged Navier--Stokes simulations (Ling {\it et al}~2016, Duraisamy {\it et al}~2019) and large-eddy simulations (Maulik and San 2017, Lozano-Dur\'an and Bae 2023) has been well examined.
As machine learning is good at nonlinearly gaining the relationship between input and output data, its usages are also seen for inverse problems including sparse sensor reconstruction and super-resolution analysis (Fukami {\it et al} 2021b, 2023).
Furthermore, machine learning enables revisiting classical problems such as identifying dominant processes (Callaham {\it et al} 2021) and important regions (Cremades {\it et al} 2024) in turbulent flows, providing physical insights that go beyond what have been possible to gain with traditional linear techniques.

Among different applications, extensive efforts have been made on machine-learning-based reduced-order modeling (ML-ROM), by which the inherently high-dimensional dynamics of fluid flows are represented by lower-dimensional dynamics.
The autoencoder (Hinton and Salakhutdinov 2006), a self-supervised machine learning designed to output the same data as the input, is a widely-used architecture for such ML-ROM.
As the autoencoder possesses a bottleneck whose data dimension is much lower than the given input, the resulting vector extracted from the latent space can be regarded as a low-order representation if the model successfully replicates the data.
Particularly for fluid flow data, it is beneficial to adopt convolutional neural networks (CNN; LeCun {\it et al} 1998) for the encoder and the decoder of an autoencoder to extract the features from coherent structures while achieving better compression than linear techniques such as proper orthogonal decomposition (Lumley 1967) which projects the data onto a flat manifold (Graham and Floryan 2021).
This CNN-based autoencoder produces the compressed representation of fluid flows as a latent vector that retains the core of flow physics residing in the high-dimensional original space (De Jes{\'u}s and Graham 2023, Page {\it et al} 2021 2024).
Furthermore, once the high-dimensional fluid flow data are compressed using the CNN-AE, the compressed low-order representation can be utilized for interpretation of fundamental physics (Omata and Shirayama 2019, Scherding {\it et al} 2023, Mo {\it et al} 2024) and design of efficient control methods (Fukagata 2023).

This review introduces the fundamentals of flow field compression using CNN-AE and its applications to various fluid dynamics problems.
The remainder of the paper is organized as follows.
In section 2, we first explain the structure and the working principle of CNN-AE with a concrete example of unsteady flows.
We also discuss the theoretical similarity and difference between the proper orthogonal decomposition and nonlinear autoencoder. 
In section 3, we cover some applications of CNN-AEs to different flow problems such as mode decomposition, latent modeling, and flow control.
Finally, a summary and an outlook are provided in section~4.

}


\section{Convolutional neural network-based autoencoder (CNN-AE)}

This section introduces a convolutional neural network-based autoencoder (CNN-AE) used for unsteady flow data compression.
A general concept of the autoencoder is provided in section~\ref{sec:AEg}.
Machine-learning models often used for CNN-AE, i.e., convolutional neural network (CNN; LeCun {\it et al} 1998) and multi-layer perceptron (MLP; Rumelhart {\it et al} 1986), are respectively detailed in sections~\ref{sec:CNN} and~\ref{sec:MLP}.
While some tips in constructing CNN-AE are offered in section~\ref{sec:CNN-AE_in_practice}, we briefly discuss the mathematical relationship between nonlinear autoencoder and linear compression techniques in section~\ref{sec:POD_vs_AE}.

\begin{figure}[b]
  \centerline{\includegraphics[clip,width=\linewidth]{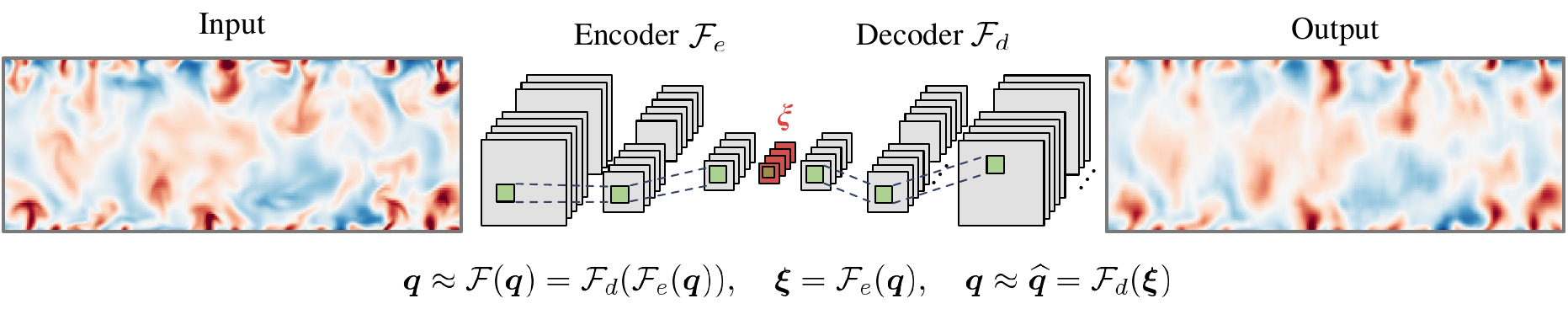}}
  \caption{
  {
  Schematic of a generic autoencoder.}
  }
  \label{fig1}
\end{figure}

\subsection{Autoencoder}
\label{sec:AEg}

Let us begin with the introduction of a general concept of autoencoder.
An autoencoder~(Hinton and Salakhutdinov 2006) is a self-supervised network trained to output $\hat{\bm{q}} \in \mathbb{R}^{N}$ from the input $\bm{q} \in \mathbb{R}^{N}$ such that $\hat{\bm{q}} \simeq \bm{q}$, where $N$ denotes the dimension of the input and output data, as illustrated in figure~\ref{fig1}.
In fluid mechanics problems, $\bm{q}$ typically represents the flow field quantities such as velocity, pressure, or vorticity fields.

A basic autoencoder consists of an encoder ${\cal F}_e$, which projects the input $\bm{q}$ into a low-dimensional latent space (or onto a manifold surface),  
\begin{equation}
\bm{\xi} = {\cal F}_e(\bm{q}),
\end{equation}
and a decoder ${\cal F}_d$,
\begin{equation}
\hat{\bm{q}} = {\cal F}_d(\bm{\xi}),
\end{equation}
which maps the latent vector $\bm{\xi} \in \mathbb{R}^{n_{\bm{\xi}}}$ back to the original dimension of the output $\hat{\bm{q}}$.
This network would be boring if $n_{\bm{\xi}}=N$, resulting in the identity mapping, but an autoencoder is usually designed with $n_{\bm{\xi}}\ll N$.
Therefore, if $\hat{\bm{q}} \simeq \bm{q}$ is attained by the training of autoencoder, it can be argued that the features of the given input data $\bm{q}$ are well represented by the low-dimensional latent vector $\bm{\xi}$.

Training of an autoencoder is performed by optimizing the weights and bias in the model.
The mean squared error between $\hat{\bm{q}}$ and  $\bm{q}$ is often considered as the loss function,
\begin{equation}
\bm{W}^* = {\rm argmin}_{\bm{W}} || \bm{q} - \bm{\hat{q}}||_2 , \label{eq:MSE}
\end{equation}
where $\bm{W}$ denotes all the weights and biases contained in the autoencoder network, and the superscript of $*$ denotes the optimum values.

This study mainly exemplifies autoencoders consisting of convolutional neural networks (CNN) and multi-layer perceptron (MLP).
However, we note that any machine learning (or non-machine learning) method can be considered as an encoder as long as it can low-dimensionalize the input data; and the same for a decoder.
For example, a combination of the Isomap-based encoder~(Tenenbaum {\it et al} 2000) and the $k$-nearest-neighbor-based decoder~(Fix and Hodges 1989) is considered for a range of controlled scenarios of fluidic pinball flows~(Marra {\it et al} 2024).
A vision transformer (Dosovitskiy {\it et al} 2000) encoder combined with a CNN decoder is also considered to extract low-dimensional features of fluid flows~(Fan {\it et al} 2025).

\begin{figure}[b]
  \centerline{\includegraphics[clip,width=0.85\linewidth]{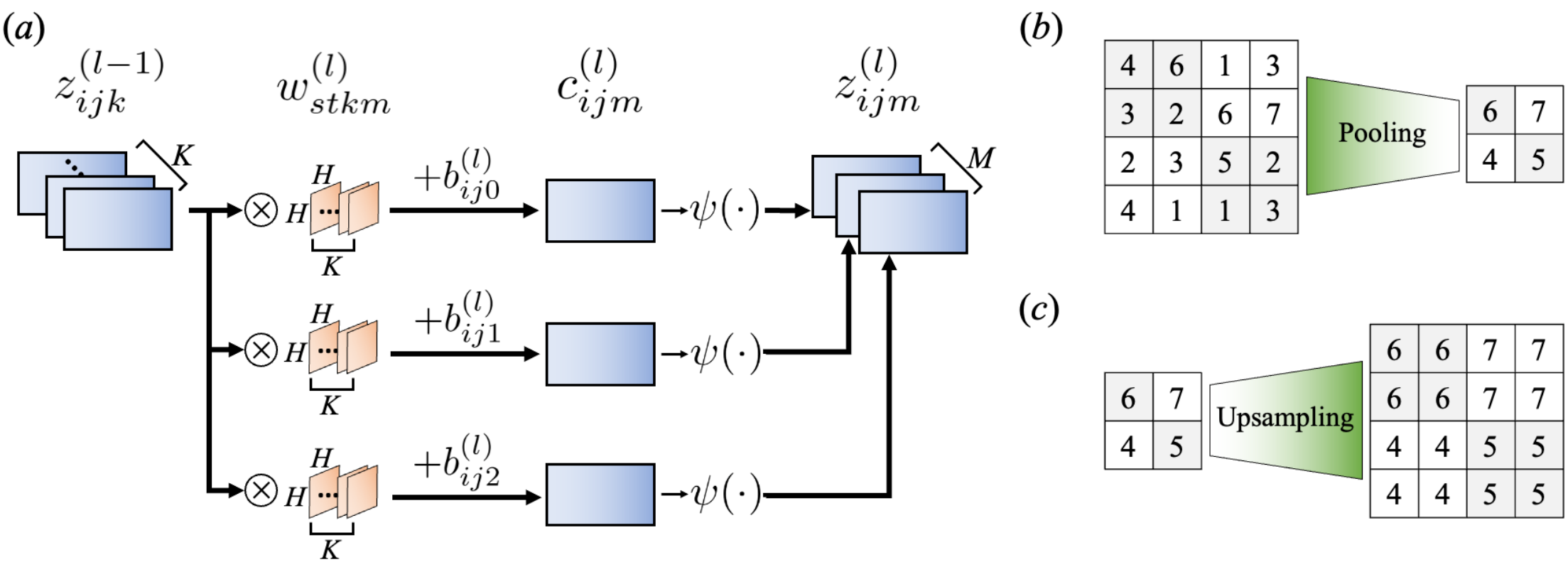}}
  \caption{Internal operations of convolutional neural network: $(a)$ convolutional layer, $(b)$ pooling layer, and $(c)$ upsampling layer. 
(Adapted from: Murata T, Fukami K and Fukagata K 2020 {\it J. Fluid Mech.}~{\bf 882} A13. 
Copyright \copyright~2019 Cambridge University Press. With permission.)
}
  \label{fig2}
\end{figure}

\subsection{Convolutional neural network (CNN)}
\label{sec:CNN}

In the CNN-AE formulation, the flow field data are first fed into the CNN portion as shown in figure~\ref{fig1} to learn large-scale structures in a flow field through filter-based operations.
The primary internal operations in CNN are illustrated in figure~\ref{fig2}.
As shown in figure~\ref{fig2}($a$), the input to the $l$th CNN layer, which is the output of the right upstream layer $z^{(l-1)}$, is convolved with the filter $w_{stkm}$ followed by an addition of bias $b_{ijm}$, i.e., 
\begin{equation}
c_{ijm}^{(l)}=\sum^{K-1}_{k=0}\sum^{H-1}_{p=0}\sum^{H-1}_{q=0} w_{stkm}^{(l)} z^{(l-1)}_{i+s-G,j+t-G,k}+b_{ijm},
\end{equation}
where $G={\rm floor} (H/2)$.
The activation function $\varphi$ is then applied to this $c_{ijm}^{(l)}$, i.e., 
\begin{equation}
z_{ijm}^{(l)}=\varphi (c_{ijm}^{(l)}),
\end{equation}
which is the output of the $l$th CNN layer.
Nonlinear functions are generally chosen for the activation function $\varphi$.
Particularly for autoencoders, it is essential to use nonlinear activation functions to achieve better compression than linear techniques, which will be discussed in section~\ref{sec:POD_vs_AE}.

In addition to the CNN layer, the pooling and upsampling layers shown in figures~\ref{fig2}($b$,~$c$) play a pivotal role in CNN-AE.
The pooling layers reduce the dimension of the data.
For the pooling methods, one can consider performing the max pooling, where the maximum value is used as illustrated in figure~\ref{fig2}($b$), or the average pooling, where the average value is instead used.
In contrast, the upsampling layers expand the dimension of data.
The simplest upsampling method is to copy the data as illustrated in figure~\ref{fig2}($c$), which is also called the nearest neighbor interpolation.
While different methods such as bilinear interpolation, max unpooling, and transverse convolution can be used, machine-learning libraries usually offer these options so that users can easily switch between them.

\subsection{Multi-layer perceptron (MLP)}
\label{sec:MLP}

Multi-layer perceptron (MLP), a conventional fully-connected neural network, is often used in the bottleneck part of the autoencoder, where the data dimension is very low and the spatial coherence is less important than the complex relationship among the components.
In a single layer of MLP, a weighted sum of the input $c^{(l)}_{i}$ is calculated, then an activation function $\varphi$ is applied,
\begin{equation}
    c^{(l)}_{i} = \sum_{j} w^{(l)}_{ij} z^{(l-1)}_{j}+b_i^{(l-1)},
    \label{eq:MLP}
\end{equation}
\begin{equation}
    z^{(l)}_{i} = \varphi(c^{(l)}_{i}).
\end{equation}
Similarly to CNN, weights $w^{(l)}_{ij}$ and bias $b^{(l-1)}_{i}$ are optimized through the training process.

\subsection{A CNN-AE in practice}
\label{sec:CNN-AE_in_practice}

This subsection offers details on the construction of a basic CNN-AE with a sample code (\url{https://github.com/kfukami/CNNAE_Practice}).
From the link above, an example routine for constructing and training the CNN-AE model shown in table~\ref{tab1} is provided with a sample data set of laminar cylinder wake at ${\rm Re}_D=100$~(Kor {\it et al} 2017, Murata {\it et al} 2020), where $D$ is a diameter of the cylinder.
Here, let us consider an input (or an output) of a two-dimensional velocity field ${\bm q} = (u,v)$ whose size is $(N_x,N_y,N_\phi)=(384,192,2)$, where $N_\phi$ describes the number of variables of interest.
For example, this $N_\phi$ can be 1 if a two-dimensional vorticity field $\omega_z$ is only used or 3 for the case in which a pressure field $p$ is additionally considered such that ${\bm q} = (u,v,p)$.

\begin{table*}[b!] 
    \caption{
    An example network structure of a basic CNN-AE with the latent dimension $n_{\bm \xi}=2$.}
    \begin{center}
    \begin{tabular}{cccc}
        \hline
        \multicolumn{2}{c}{Encoder} & \multicolumn{2}{c}{Decoder} \\ \hline
        Layer & Data size  & Layer & Data size \\ \hline
        Input & $(384,192,2)$ & Latent vector & $(2,1,1)$  \\
        1st Conv. $(3,3,16)$ & $(384,192,16)$ & Fully connected & $(6,3,4)$ \\
        1st MaxPooling & $(192,96,16)$ & 1st Upsampling & $(12,6,4)$\\
        2nd Conv. $(3,3,8)$ & $(192,96,8)$ & 7th Conv. $(3,3,4)$ & $(12,6,4)$\\
        2nd MaxPooling & $(96,48,8)$ & 2nd Upsampling & $(24,12,4)$\\
        3rd Conv. $(3,3,8)$ & $(96,48,8)$ & 8th Conv. $(3,3,8)$ & $(24,12,8)$\\
        3rd MaxPooling & $(48,24,8)$ & 3rd Upsampling & $(48,24,8)$\\
        4th Conv. $(3,3,8)$ & $(48,24,8)$ & 9th Conv. $(3,3,8)$ & $(48,24,8)$\\
        4th MaxPooling & $(24,12,8)$ & 4th Upsampling & $(96,48,8)$ \\
        5th Conv. $(3,3,4)$ & $(24,12,4)$ & 10th Conv. $(3,3,8)$ & $(96,48,8)$\\
        5th MaxPooling & $(12,6,4)$ & 5th Upsampling & $(192,96,8)$\\
        6th Conv. $(3,3,4)$ & $(12,6,4)$ & 11th Conv. $(3,3,16)$ & $(192,96,16)$\\
        6th MaxPooling & $(6,3,4)$ & 6th Upsampling & $(384,192,16)$\\
        Fully connected & $(2,1,1)$ & 12th Conv. $(3,3,2)$ & $(384,192,2)$\\
        (Latent vector) & & (Output)  &  \\
        \hline
    \end{tabular}
    \end{center} \label{tab1}
    \vspace{-3mm}
\end{table*}

The input data are then gradually compressed through convolutional and maxpooling operations.
The example model above uses a filter (or kernel) size of $3^2$.
While the kernel size is one of the hyperparameters, a range of filter sizes (e.g., 3, 5, 7, and 9) can also be considered across the layer direction to capture multi-scale features of unsteady flows~(Fukami {\it et al} 2024a).
Note that an odd number is generally used for the filter size to avoid asymmetric filtering operations inside a CNN-AE~(Morimoto {\it et al} 2021).

The pooling and upsampling ratio are set to $2^2$ in the code --- this can also vary depending on the size of the given data. 
However, excessive compression and expansion at once may not be preferable as the hidden layer would necessitate the nonlinear relationship between the layers possessing a large difference in the degree of freedom. 
Rather, gradual compression such as the ratio of $2^2$ enhances giving a physically interpretable role to each of the hidden layers in a nonlinear machine-learning model.
It is known that such a convolutional model often captures large-scale structures of fluid flows at earlier layers, while local, fine-scale features are learnt in a deeper layer~(Motimoto {\it et al} 2022, Guan {\it et al} 2022).
Since the pooling or upsampling layer is solely used for data-size adjustment and is not trainable, a careful choice of the ratio is essential to enable efficient nonlinear compression of fluid flows through a CNN-AE.

While the example model here includes only a single layer of MLP between the convolutional layers and the latent vector, one can consider adding more hidden MLP layers to further enhance compression capability of the autoencoder model.
The balance between the CNN and MLP parts is one hyperparameter in constructing the autoencoder for fluid flows.
More detailed discussions on this point of CNN-MLP balance as well as the influence of nonlinear activation functions are offered in Fukami {\it et al} (2021a).

\subsection{Comparison between a linear POD and nonlinear AE}
\label{sec:POD_vs_AE}

Proper orthogonal decomposition (POD; Lumley 1967) is one of the most widely used linear modal decomposition methods in the fluid mechanics community. 
The essence of the POD is to identify the set of orthogonal modes, i.e.,  bases, by which the given data are expressed by the least number of modes.
Therefore, one can reduce the data dimension by truncating the modes whose contributions are small.
Among different formulations of POD, the snapshot POD is often used for fluid flow data due to its lower computational cost (Sirovich 1987).
Here, we skip further details of POD; readers are referred to, e.g., Taira {\it et al} (2017).
Rather, we here briefly compare the POD and nonlinear autoencoder in terms of their mathematical formulations.

\begin{figure}[b]
    \centering
    \includegraphics[width=1\textwidth]{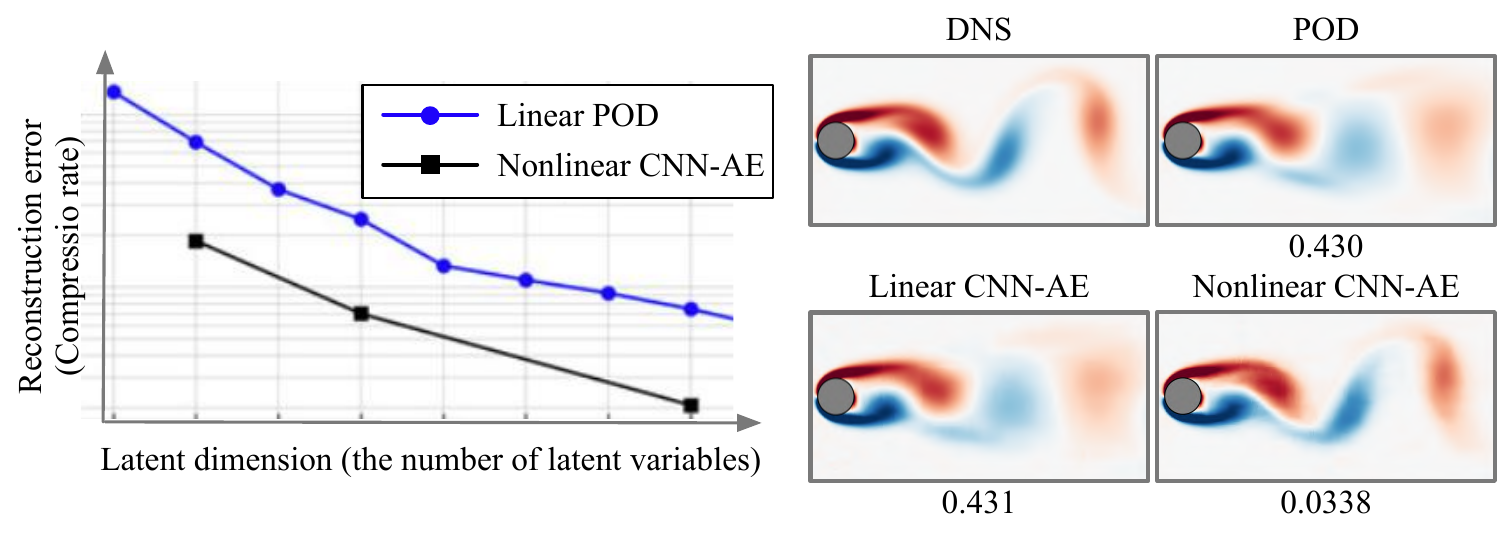}
    \vspace{-5mm}
    \caption{
    Linear and nonlinear compression of fluid flows.
    Nonlinear models generally provide better compression than linear models due to the use of nonlinear functions inside their frameworks such as activation functions and kernels.
    Example reconstruction fields for laminar cylinder wake at ${\rm Re}_D=100$ using POD, a linear {CNN-AE,}
    and a nonlinear {CNN-AE}
    with $n_{\bm \xi}$ are also shown.
    The value underneath each contour reports the $L_2$ error norm.
    }
    \label{fig_AEgene}
\end{figure}

Along with the reduced basis $\bm{\Gamma}\in \mathbb{R}^{n_{\bm{\xi}}\times N}$, the POD is expressed as an $L_2$ minimization problem such as
\begin{equation}
\bm{\Gamma}^* = {\rm argmin}_{\bm{\Gamma}} ||  \bm{q}^\prime - \bm{\Gamma}^{\rm T} \bm{\Gamma} \bm{q}^\prime ||_2 ,
\end{equation}
where $\bm{q}^\prime = (\bm{q} - \overline{\bm{q}}) \in \mathbb{R}^{n_t \times n_{\bm{\xi}}}$ denotes the collection of snapshots of the fluctuating components.
In contrast, the training of a nonlinear autoencoder is expressed in a similar manner,
\begin{equation}
\bm{\Gamma}^* = {\rm argmin}_{\bm{\Gamma}} ||  \bm{q}^\prime - \varphi_d \bm{\Gamma}^{\rm T} \varphi_e \bm{\Gamma} \bm{q}^\prime ||_2 ,
\end{equation}
where $\bm{\Gamma}$ is the summation weights in the network, and $\varphi_e$ and $\varphi_d$ represent the activation functions of the encoder and decoder, respectively.
These two equations suggest that nonlinear autoencoder is similar to POD --- in particular, these two are identical when the activation function is linear, i.e., $\varphi_e=\varphi_d=1$ (Baldi and Hornik 1989, Milano and Koumoutsakos 2002, Murata {\it et al} 2020, Fukami {\it et al} 2021a).
The use of nonlinear activation functions enables the reduction of reconstruction error at the same number of modes $n_{\bm{\xi}}$ compared to linear techniques.

The above points are evident from a comparison between data compressions using POD, a linear CNN-AE, and a nonlinear CNN-AE similar to what we introduced in section~\ref{sec:CNN-AE_in_practice}, for a laminar cylinder wake at ${\rm Re}_D=100$ presented in figure~\ref{fig_AEgene}.
The reconstruction error of the nonlinear autoencoder is always smaller than that of the POD and the linear CNN-AE.
The vorticity fields reconstructed by the POD and the linear autoencoder are quite similar and the wake structures are substantially diffused as compared to the ground truth computed by DNS.
In contrast, the nonlinear CNN-AE successfully captures the wake structure.
This example demonstrates that nonlinear autoencoders better compress the fluid flow data than conventional linear methods.

\begin{figure}[b]
    \centering
    \includegraphics[width=1\textwidth]{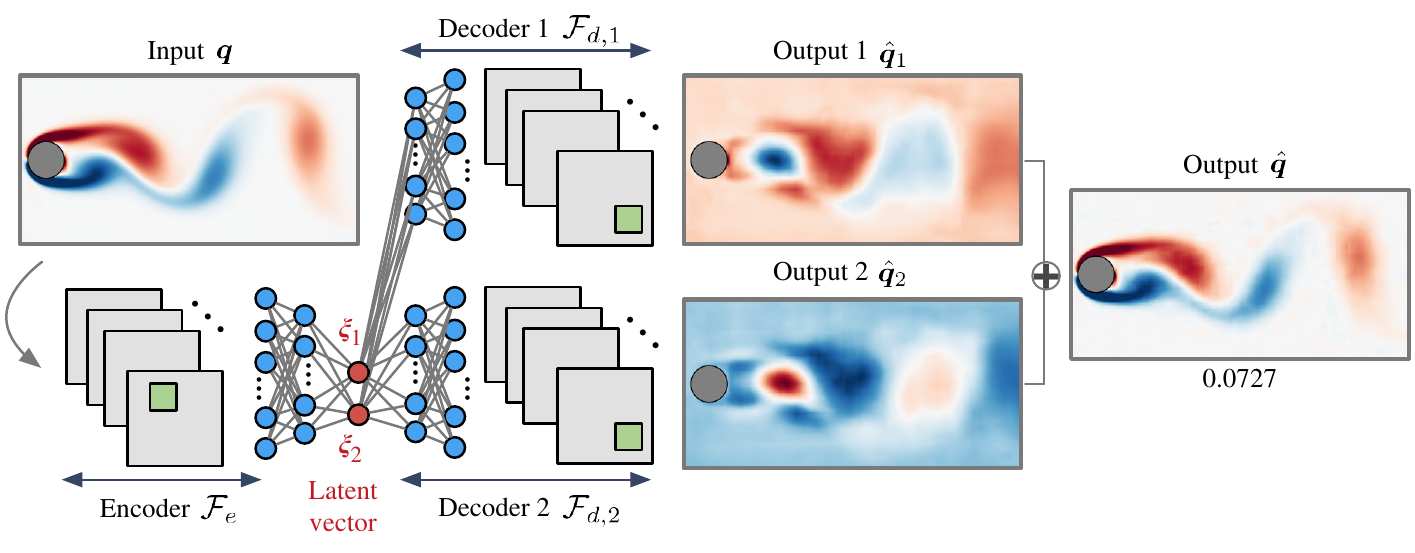}
    \vspace{-5mm}
    \caption{Mode-decomposing 
    {CNN-AE}
    (Murata {\it et al} 2020).}
    \label{fig_MDCNNAE}
\end{figure}

\section{Applications of CNN-AEs}

In this section, we discuss some representative applications of CNN-AEs.
The applications covered herein are categorized into mode decomposition (section~\ref{sec:MD}), latent modeling (section~\ref{sec:LM}), and flow control (section~\ref{sec:FC}).

\subsection{Mode decomposition}
\label{sec:MD}


As discussed above, one can low-dimensionalize flow fields using the basic CNN-AE in section~\ref{sec:CNN-AE_in_practice}.
However, it is often challenging to interpret the physical meaning of each latent mode because the latent variables are decoded altogether.
For better interpretation of latent modes, Murata {\it et al} (2020) proposed a mode-decomposing CNN-AE (MD-CNN-AE), as illustrated in figure~\ref{fig_MDCNNAE}.
While the decoder part of the MD-CNN-AE is the same as that of the basic CNN-AE, the decoders are designed to separately decode each latent variable,
\begin{equation}
    \hat{\bm q}_1 = {\cal F}_{d,1}({\xi_1}),~~~\hat{\bm q}_2 = {\cal F}_{d,2}({\xi_2}), ~~~\hat{\bm q} = \hat{\bm q}_1 + \hat{\bm q}_2, 
\end{equation}
such that the role of each latent variable becomes clearer.

\begin{figure}
    \centering
    \includegraphics[width=0.85\textwidth]{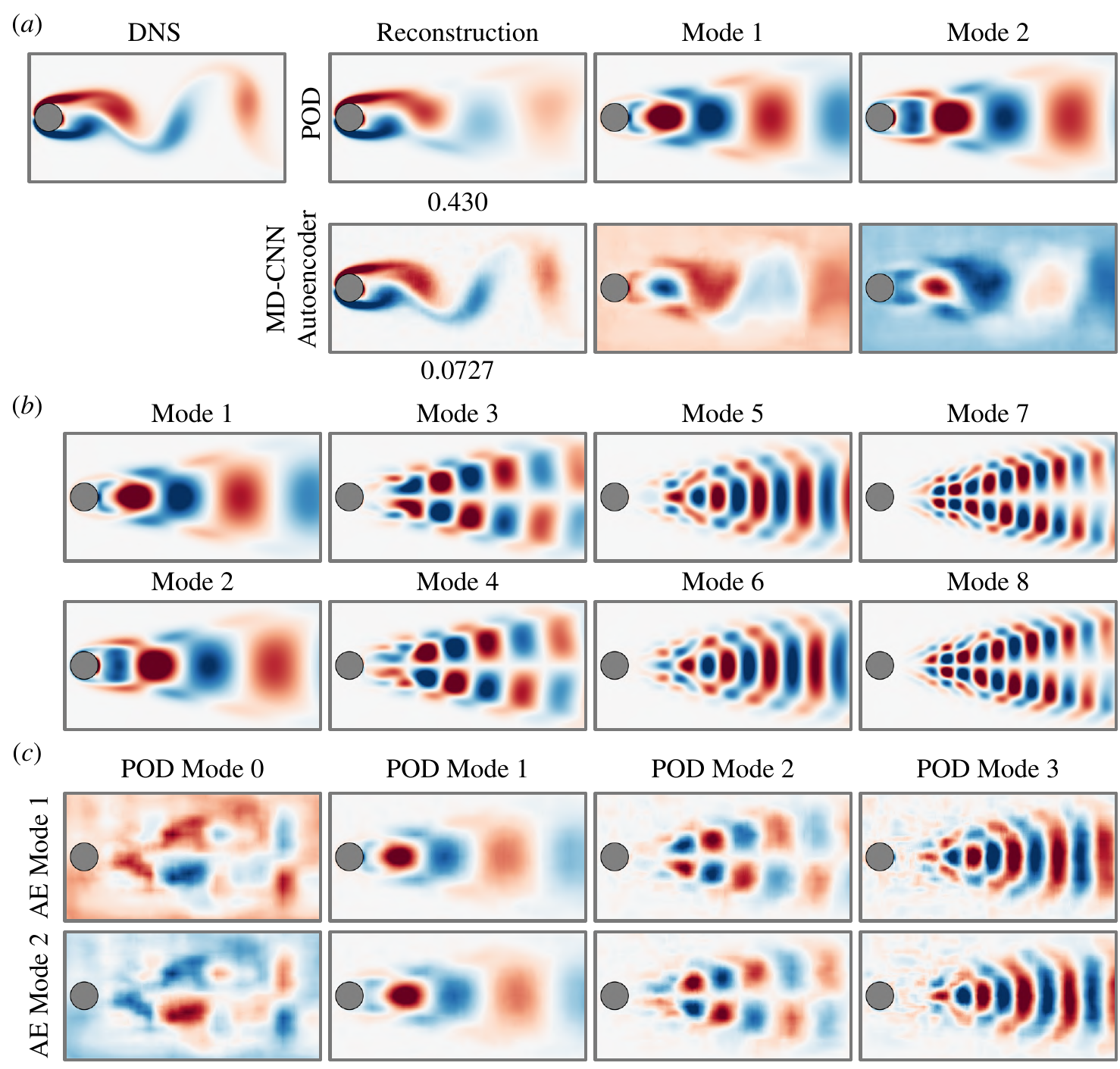}
    \caption{
    Nonlinear machine-learning-based compression of laminar cylinder wake.
    $(a)$~Decoded field and modes of POD and MD-CNN-AE with $n_{\bm \xi}=2$.
    The value underneath each reconstructed field reports the $L_2$ error norm.
    $(b)$~The dominant eight POD modes of cylinder wake.
    $(c)$~POD modes extracted from nonlinear-autoencoder modes.
    (Adapted from:
    Fukami K, Nakamura T and Fukagata K 2020
    {\it Phys.~Fluids} {\bf 32} 095110.
    Copyright \copyright~2020 Authors.)   
    }
    \label{fig_MDAEmodes}
\end{figure}

We compare mode decomposition through POD and MD-CNN-AE for a two-dimensional flow around a circular cylinder at the Reynolds number ${\rm Re}_D=100$ (Murata {\it et al} 2020, Fukami {\it et al} 2020b) in figure~\ref{fig_MDAEmodes}.
The instantaneous reconstructed vorticity field and modes with $n_{\bm \xi}=2$ are presented in figure~\ref{fig_MDAEmodes}$(a)$.
Similar to the observation in figure~\ref{fig_AEgene}, the vorticity field reconstructed by the MD-CNN-AE much resembles that of DNS, which is also quantitatively supported by the $L_2$ error norm presented underneath each subfigure.

This superior compression of nonlinear MD-CNN-AE can be examined by performing POD for the collection of nonlinear-autoencoder modes, as depicted in figures~\ref{fig_MDAEmodes}$(b)$ and $(c)$.
Here, we showcase the dominant eight POD modes of cylinder wake and the POD modes extracted from the collection of nonlinear-autoencoder modes, respectively.
The POD mode 1 extracted from the nonlinear modes 1 and 2 is similar to modes 1 and 2 of the original POD, the POD mode 2 extracted from the nonlinear modes 1 and 2 is similar to modes 3 and 4 of the original POD, and the POD mode 3 extracted from the nonlinear modes 1 and 2 is similar to modes 5 and 6 of the original POD.
Although the POD mode 0 extracted from the nonlinear modes 1 and 2 does not show a clear structure, these two simply cancel with each other (Murata {\it et al} 2020).
The observation above suggests that each nonlinear-autoencoder mode contains multiple POD modes, providing better reconstruction with the MD-CNN-AE compared to the POD under the same number of modes --- in this example of two-dimensional laminar cylinder wake, a single nonlinear-autoencoder mode includes three POD modes.

In principle, MD-CNN-AE can be used for more complex problems with higher degrees of freedom.
Ando {\it et al} (2023) applied the MD-CNN-AE to three-dimensional flow fields by performing large-scale distributed learning on Fugaku supercomputer, evaluating the applicability to flows around complex shapes of car bodies obtained by large-scale simulations.
However, if possible, a method with even higher compression rate is desired in considering more complex flows such as turbulent flows.
It is also preferable to arrange the autoencoder modes in the order of their energy contents, which is not the case with conventional networks.
In response, Fukami {\it et al} (2020b) proposed a hierarchical CNN autoencoder (HAE) for fluid flows inspired by the idea of Saegusa {\it et al} (2004).

\begin{figure}
    \centering
    \includegraphics[width=0.9\textwidth]{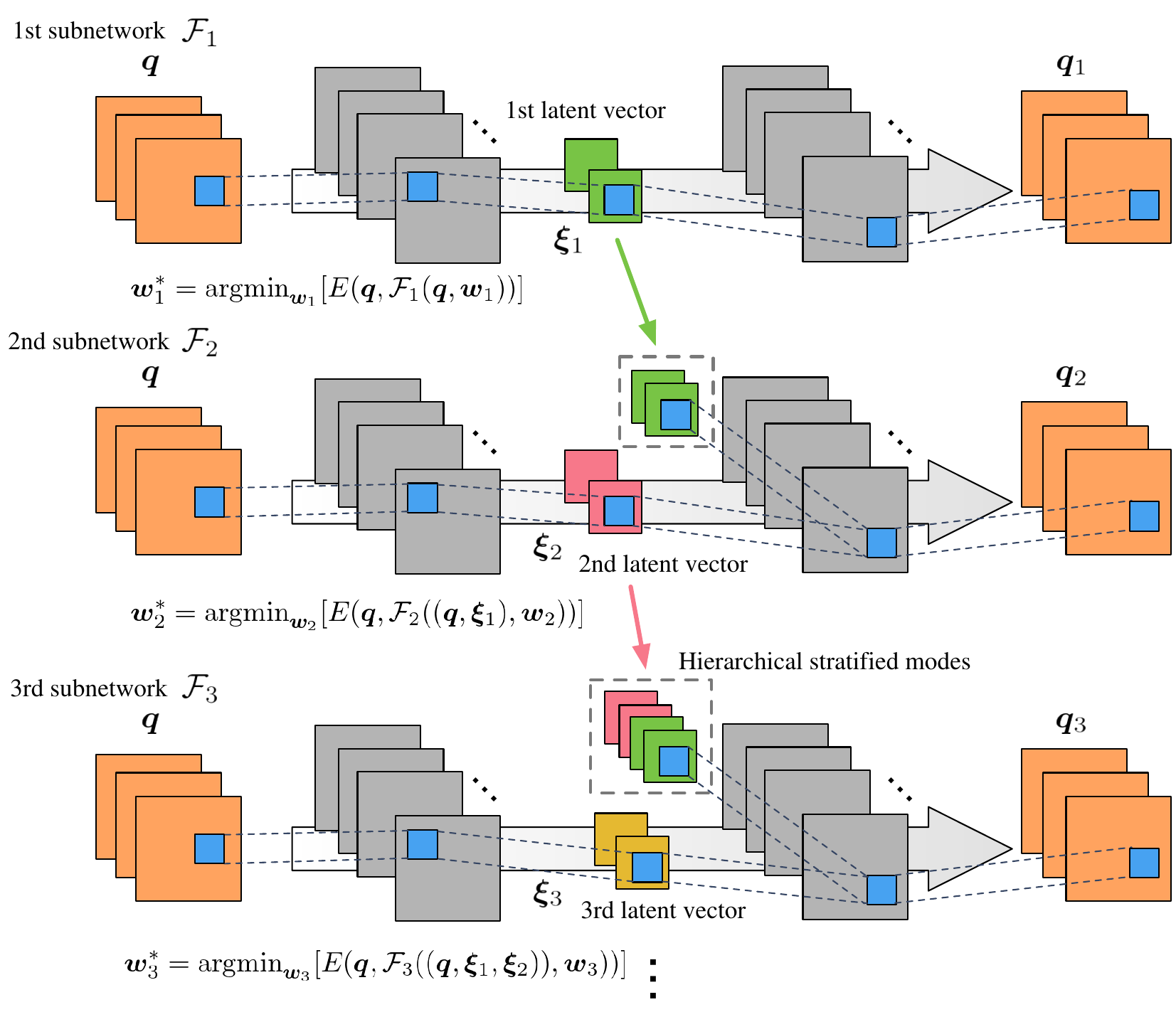}
    \caption{
    Hierarchical nonlinear autoencoder.
    (Adapted from:
    Fukami K, Nakamura T and Fukagata K 2020
    {\it Phys.~Fluids} {\bf 32} 095110.
    Copyright \copyright~2020 Authors.)   
    }
    \label{fig_HAE}
\end{figure}

The network structure of HAE (Fukami {\it et al} 2020b) is illustrated in figure~\ref{fig_HAE}.
As its name implies, training of the HAE is performed in a hierarchical manner.
The 1st subnetwork ${\cal F}_1$ is trained similarly to that for the basic CNN-AE to obtain the 1st latent vector $\bm{\xi}_1$.
The obtained 1st latent vector $\bm{\xi}_1$ is concatenated to the bottleneck part of the 2nd subnetwork ${\cal F}_2$.
This ${\cal F}_2$ is trained while fixing $\bm{\xi}_1$ to obtain the 2nd latent vector $\bm{\xi}_2$.
In other words, the second latent vector $\bm{\xi}_2$ can be regarded as low-dimensionalized residual information of the flows that cannot be extracted from the first network.
Repeatedly performing this hierarchical training, nonlinear CNN-AE modes can be obtained while ranking them with respect to their energy contents.
These procedures are mathematically described as,
\begin{eqnarray}
    {{\bm q}_1} = {\cal F}_1({\bm q};{\bm w}_1)={\cal F}_{d,1}({\bm \xi}_1),~  {{\bm\xi}_1}={\cal F}_{e,1}(\bm q),\nonumber\\
    {{\bm q}_2}= {\cal F}_2({\bm q},{{\bm \xi}_1};{\bm w}_2)= {\cal F}_{d,2}(\{{\bm \xi}_1,{\bm \xi}_2\}),~ 
    {{\bm\xi}_2}={\cal F}_{e,2}(\bm x),\nonumber\\
    {{\bm q}_3}= {\cal F}_3({\bm q}, \{{{\bm \xi}_1},{{\bm \xi}_2} \};{\bm w}_3 )={\cal F}_{d,3}(\{{\bm \xi}_1,{\bm \xi}_2,{\bm \xi}_3\}),~  {{\bm\xi}_3}={\cal F}_{e,3}(\bm q),\nonumber\\
    \vdots\nonumber \\
    {{\bm q}_M}= {\cal F}_M({\bm q}, \{ {{\bm \xi}_1},\dots,{{\bm \xi}_{M-1}} \};{\bm w}_M)={\cal F}_{d,M}(\{ {{\bm \xi}_1},\dots,{{\bm \xi}_M} \}),\nonumber\\
    {{\bm\xi}_M}={\cal F}_{e,M}(\bm q),
\end{eqnarray}
where $M$ is the number of the contained autoencoder modes.

Note that each latent vector ${\bm \xi}_i$ in the HAE formulation above can be either a scalar or a vector such that multiple nonlinear-autoencoder modes can be involved for each subnetwork.  
The latter is particularly useful in considering flows necessitating a large number of modes such as turbulence.
In such cases, each ${\bm \xi}_i$ is expressed as a {\it mode family} (Fukami {\it et al} 2020b).
Although individual members (scalar modes) in each mode family are not ranked, the mode families are globally ordered following their energy contents.
In what follows, the former of a scalar case is applied to a laminar cylinder wake while the latter of a vector scenario is considered for turbulent channel flow.

\begin{figure}
    \centering
    \includegraphics[width=0.9\textwidth]{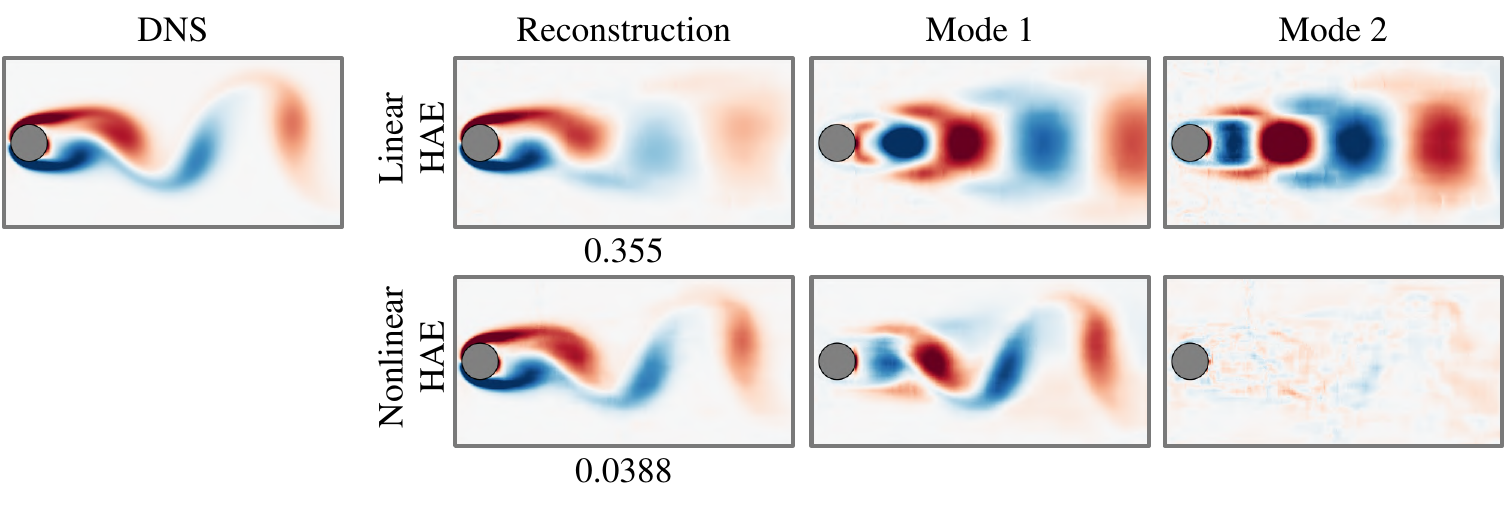}
    \caption{
    Decoded field and modes of POD and hierarchical nonlinear autoencoder with $n_{\bm \xi}=2$ for laminar cylinder wake.
    The value underneath each reconstructed field reports the $L_2$ error norm.
    (Adapted from:
    Fukami K, Nakamura T and Fukagata K 2020
    {\it Phys.~Fluids} {\bf 32} 095110.
    Copyright \copyright~2020 Authors.)   
    }
    \label{fig_HAE_Rec}
\end{figure}

The decoded field and modes of linear and nonlinear hierarchical nonlinear autoencoder (HAE) with $n_{\bm \xi}=2$ for laminar cylinder wake are presented in figure~\ref{fig_HAE_Rec} (Fukami {\it et al} 2020b).
The linear HAE provides almost identical reconstruction and modes to POD, suggesting that the hierarchical formulation successfully produces the ranked modes even from the network-based training.
The flow field data are mostly compressed into mode 1 of HAE while mode 2 of HAE exhibits almost zero.
This implies that there is almost no residual for mode 2 if we consider nonlinear activation functions.
In other words, a laminar cylinder wake can be compressed into solely a single scaler, which coincides with studies for phase-based reduced-order models reporting that a time-varying flow state can be projected onto a single phase variable $\theta$ over a wake-shedding period (Taira and Nakao 2018).
HAE modes also contain eight POD modes of laminar cylinder wake as well as the MD-CNN-AE modes, although not shown here.
Note that mode 2 of HAE can exhibit interpretable residual information when applying more complex cases such as transient flow around a cylinder (Fukami {\it et al} 2020b).

\begin{figure}
    \centering
    \includegraphics[width=\textwidth]{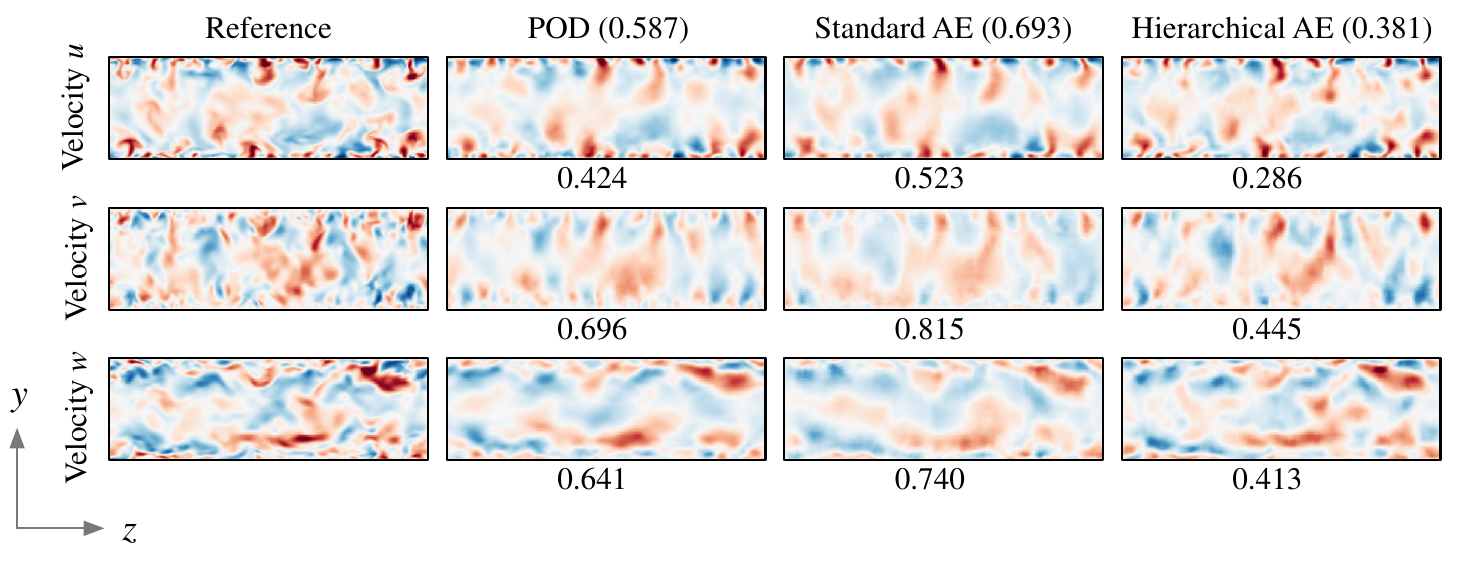}
    \caption{
     Reconstructed velocity fields from POD, a standard autoencoder, and the hierarchical autoencoder with $n_{\bm \xi} = 288$ of turbulent channel flow. 
     The value underneath each contour reports the $L_2$ error norm.
    (Adapted from:
    Fukami K, Nakamura T and Fukagata K 2020
    {\it Phys.~Fluids} {\bf 32} 095110.
    Copyright \copyright~2020 Authors.)      
    }
    \label{fig_HAE_Channel}
\end{figure}

This hierarchical training provides better compression than its regular counterpart, particularly for more complex problems with a much higher degree of freedom.
Let us apply the HAE to a turbulent channel flow at the friction Reynolds number $Re_\tau=180$.
The instantaneous cross-sectional velocity fields reconstructed using POD, a standard CNN-AE, and the HAE with $n_{\bm \xi} = 288$ are compared in figure~\ref{fig_HAE_Channel}. 
As mentioned above, each hierarchical mode for this example of turbulent channel flow includes multiple nonlinear modes as a mode family, 72 modes/family $\times$ 4 modes, due to a large degree of freedom.

While POD and a standard nonlinear autoencoder reconstruct large-scale structures only, the HAE reproduces finer structures compared to them, which is also evident from the $L_2$ reconstruction error norm for all three velocity components.
Although it may be challenging to interpret the nonlinear modes with such a case involving mode families, the hierarchical training, somewhat similar to transfer learning~(Inubushi and Goto 2020, Guastoni {\it et al} 2021), achieves better compression of unsteady flow data than other techniques.

One shortcoming of such a transfer learning-based idea is the need for a fine-tuning process.
The HAE exactly belongs to this case since the number of decoders increases with the latent dimension.
In other words, expensive neural network-training is necessitated repeatedly, which is not ideal, particularly for cases with excessively large degrees of freedom in space and time such as three-dimensional turbulent flows. 
In response, Eivazi {\it et al} (2022) considered a $\beta$-variational autoencoder-based compression of unsteady flows, providing the ranked CNN-AE modes with a one-shot training.

\begin{figure}[t]
    \centering
    \includegraphics[width=\textwidth]{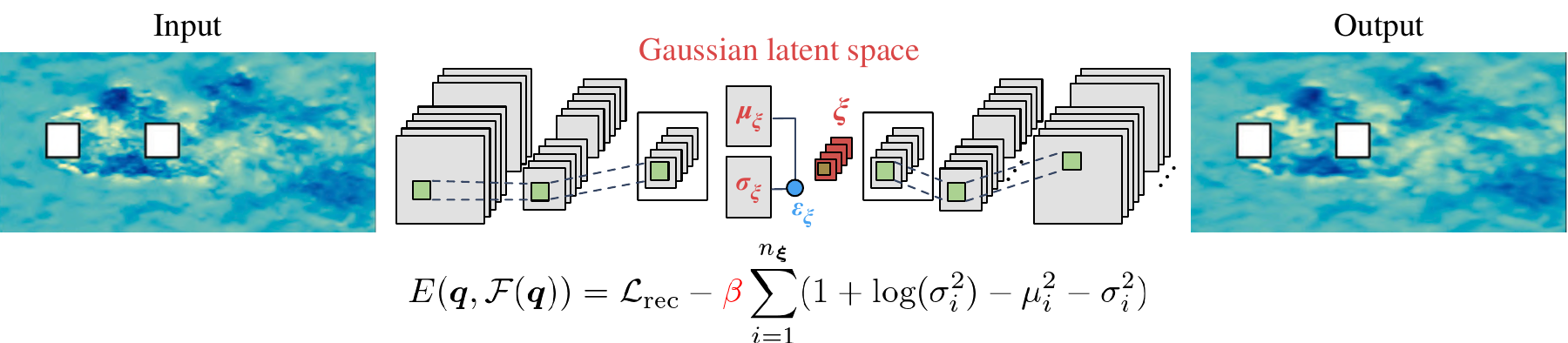}
    \caption{
    The $\beta$-variational autoencoder.
    (Input and output images are adapted from: 
    Eivazi H, Le Clainche S, Hoyas S and Vinuesa R 2022
    {\it Expert Syst.~Appl.}~{\bf 202} 117038.
    Copyright \copyright~2022 The Authors, CC BY 4.0 license.)
    }
    \label{fig_betaVAE}
\end{figure}

The main difference of $\beta$-variational autoencoder from a standard model is its latent space construction, as illustrated in figure~\ref{fig_betaVAE}.
The latent variables here are parameterized in a Gaussian manner such that
\begin{eqnarray}
    {\bm \xi}\sim {\cal N}({\bm \mu}_{\bm \xi},{\bm \sigma}_{\bm \xi}^2),
\end{eqnarray}
where ${\bm \mu}_{\bm \xi}$ and ${\bm \sigma}_{\bm \xi}$ represent the mean and standard deviation of the latent vector. 
Autoencoders enforcing this Gaussian form in the latent representation are referred to as variational autoencoders (VAE; Kingma and Welling 2013), which are often considered for generative learning in image science.
Eivazi {\it et al} (2022) focused on its variant called a $\beta$-VAE~(Higgins {\it et al} 2017), which penalizes the latent data dimension $n_{\bm \xi}$ while providing statistically-independent, nearly orthogonal latent variables.
This enables seeking the minimal representation of interpretable low-order subspace, which may be useful characteristics for low-dimensionalizing fluid flows.

The above-mentioned promotion of parsimoniousness and orthogonality is achieved by minimizing the loss function below,
\begin{eqnarray}
    E({\bm q},{\cal F}({\bm q})) = {\cal L}_{\rm rec} - \beta \sum_{i=1}^{n_{\bm \xi}} (1+\log(\sigma_i^2)-\mu_i^2 - \sigma_i^2),~\label{eq:vae}
\end{eqnarray}
where ${\cal L}_{\rm rec}$ is a regular reconstruction loss between the input and output of CNN-AE such as the mean squared error appearing in equation~(\ref{eq:MSE}).
The second term weighed with a balancing coefficient $\beta$ contributes to the parsimoniousness and orthogonality of latent representation.
As this formulation cannot provide the ranked modes that are available with the hierarchical autoencoder, the network proposed in Eivazi {\it et al} (2022) additionally involves an energy-based mode ranking algorithm. 
More details of the formulation are given in Eivazi {\it et al} (2022).

\begin{figure}
    \centering
    \includegraphics[width=0.85\textwidth]{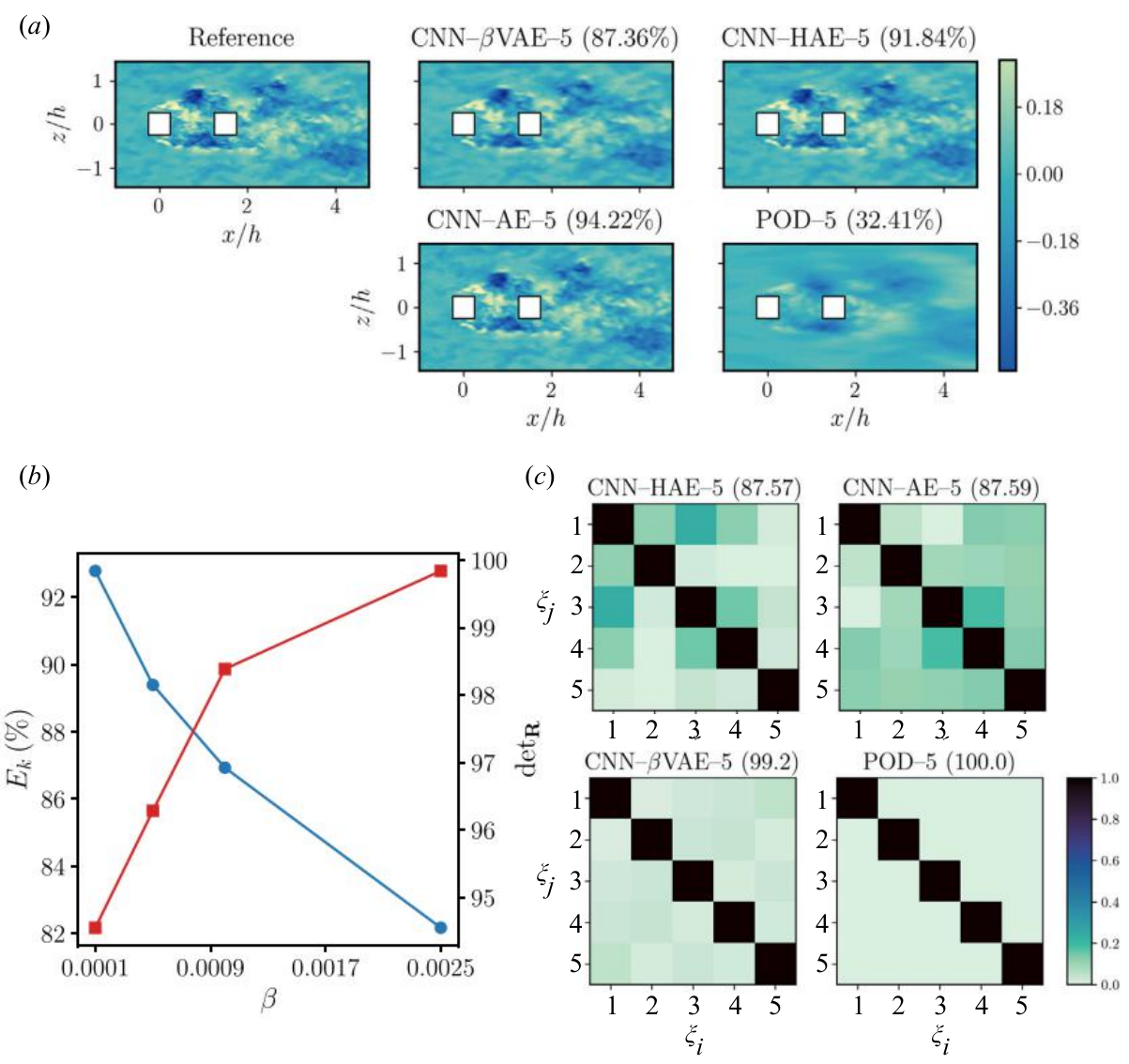}
    \caption{{
    $\beta$-variational autoencoder-based compression of unsteady flows.
    $(a)$ Reconstruction of the fluctuating component of the streamwise velocity obtained from different methods. The value in brackets above each contour indicates the reconstructed kinetic energy ratio.
    $(b)$ Dependence of the reconstructed kinetic energy ratio $E_k$ and the orthogonality quantified with ${\rm det}_{\rm{{R}}}$ on the weighting coefficient $\beta$, where ${\rm{{R}}}$ is the correlation matrix of latent variables shown in $(c)$.
    The value in brackets above each contour of figure $(c)$ reports ${\rm det}_{\rm{{R}}}$.
    (Adapted from: 
    Eivazi H, Le Clainche S, Hoyas S and Vinuesa R 2022
    {\it Expert Syst.~Appl.}~{\bf 202} 117038.
    Copyright \copyright~2022 The Authors, CC BY 4.0 license.)
    }
    }
    \label{fig_betaVAE_res}
\end{figure}

The $\beta$-variational autoencoder-based compression for an example of the flow through a simplified urban environment is performed in figure~\ref{fig_betaVAE_res} (Eivazi {\it et al} 2022).
The reconstruction by the CNN-$\beta$-VAE with $n_{\bm \xi} = 5$ and $\beta = 10^{-3}$ is compared to that by a regular CNN-AE, the hierarchical autoencoder (Fukami {\it et al} 2020b), and linear POD in figure~\ref{fig_betaVAE_res}$(a)$.
The CNN-$\beta$-VAE presents better reconstruction of flow separation and wake behavior around the bodies than POD in terms of kinetic energy ratio reported in brackets, although it is slightly worse than the other two nonlinear CNN-AEs.
This is because of the cost function in equation~(\ref{eq:vae}) which accounts for the orthogonality and latent dimension.

They also examine the dependence of the reconstructed kinetic energy ratio $E_k$ and the orthogonality on the weighting coefficient $\beta$, as shown in figure~\ref{fig_betaVAE_res}$(b)$.
Here, the statistical independence between the latent variables is quantified with ${\rm det}_{\rm{{R}}}$, where ${\rm{{R}}}$ is the correlation matrix of latent variables, as presented in figure~\ref{fig_betaVAE_res}$(c)$.
Note that the larger ${\rm det}_{\rm{{R}}}$, more orthogonal since the latent variables would be more statistically independent of each other and its value of POD (completely orthogonal) is 100.
As $\beta$ increases, the reconstruction becomes worse while its orthogonality is more ensured.
The visualization of the correlation matrix ${\rm{{R}}}$ clearly presents that the latent value distribution of the CNN-$\beta$-VAE is much closer to POD compared to the other networks, reporting a high ${\rm det}_{\rm{{R}}}$ of 99.2.
Users should be mindful of the trade-off relationship between the reconstruction error and linearity since the excessive penalty of the orthogonality simply makes nonlinear autoencoder become linear POD.
However, appropriate weighting as performed by Eivazi {\it et al} (2022) may provide interesting opportunities for nonlinear modal analysis with autoencoder by combining it with conventional techniques such as the POD Galerkin approach developed for linear, ranked orthogonal modes.

\begin{figure}
    \centering
    \includegraphics[width=\textwidth]{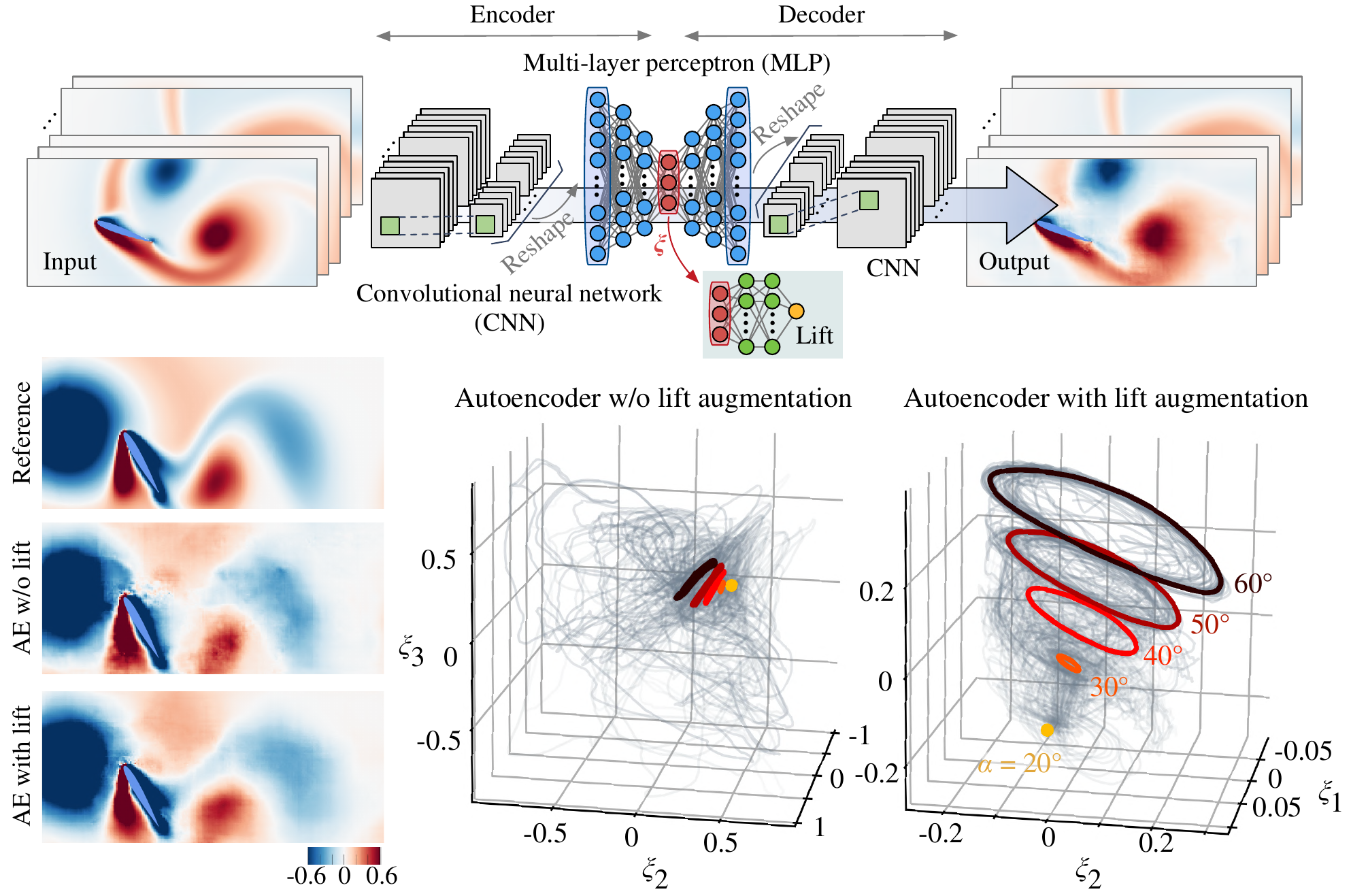}
    \caption{
    Observable-augmented nonlinear autoencoder for extreme vortex-airfoil interactions (Fukami and Taira 2023).
    The three-dimensional latent expression with the reconstructed data of a representative vorticity field is shown for a standard and the observable network. 
    (Adapted from: 
    Fukami K and Taira K 2023 {\it Nat.~Commun.}~{\bf 14} 6480.
    Copyright \copyright~2023 The Authors(s), CC BY 4.0 license.)
    }
    \label{fig_XAero_Comp}
\end{figure}

While the CNN-AEs considered above have aimed to extract nonlinear low-order modes through the modification of model structures and optimization routine, one can also consider explicitly giving physical variables in identifying a low-order subspace.
Such a physics embedding into the latent space identification can be achieved by an observable-augmented nonlinear CNN-AE proposed by Fukami and Taira (2023), as illustrated in figure~\ref{fig_XAero_Comp}.
This CNN-AE model includes an additional subnetwork that estimates a physical observable such as sensor readings and aerodynamic loads from the latent vector ${\bm \xi}$.
Hence, training of observable-augmented CNN-AE is performed by minimizing the cost function,
\begin{eqnarray}
        {\bm W}^* = {\rm argmin}_{\bm W}[||{\bm q}-\hat{\bm q}||_2 + \beta ||\kappa - \widehat{\kappa}||_2],
\end{eqnarray}
where $\kappa$ is a given observable.
Throughout this simple addition of the branch network, it is anticipated that the weights inside the CNN-AE would be tuned to learn important flow structures associated with a given physical observable in compressing the flow field data to minimize the above cost function.

The observable-augmented autoencoder is originally proposed for unsteady flows under extremely strong vortex gust-wing interactions~(Fukami and Taira 2023).
They consider a flow around a NACA0012 wing at a chord length-based Reynolds number of 100 while covering a range of angles of attack $\alpha$ with various forms of vortex gust encounter parameterized with gust size, gust strength, and its orientation against a wing.
At the Reynolds number they consider, the undisturbed baseline flow at $\alpha=20^\circ$ is steady while that for $\alpha \gtrsim 30^\circ$ presents periodic wake shedding.
The disturbed flows, on the other hand, exhibit massive separation due to extreme vortex-airfoil interactions with a large excitation of aerodynamic forces within a very short time duration.

Shown in the bottom half of figure~\ref{fig_XAero_Comp} are the reconstructed flow field and the resulting latent space with $n_{\bm \xi}=3$ via a regular CNN-AE and the observable-augmented CNN-AE.
Here, the vorticity field $\omega$ is used as the input and output of the main autoencoder network, and the lift coefficient $C_L$ is given as an observable output of the subnetwork.

Despite the fact that both autoencoders accurately reconstruct a vortical flow state regardless of the lift-based augmentation, their latent spaces present significant differences in the shape of the identified submanifold.
The trajectories emphasized in color correspond to the time series of the baseline undisturbed flow snapshots while those colored in gray are projected from the whole set of time-varying disturbed flows spanning over a huge parameter space.
For a regular CNN-AE, the latent representations of undisturbed and disturbed cases are unorganized, likely because the model focuses solely on distinguishing given cases from the parameter space by widely covering the low-order space.

In contrast, the lift-augmented CNN-AE provides a coherent geometric structure --- that is, a cone-shaped submanifold in the three-dimensional space.
For the undisturbed baseline flows, the steady fixed point of $\alpha = 20^\circ$ and the other unsteady periodic limit cycle oscillations for $\alpha \geq 30^\circ$ are expressed in a low-order, interpretable manner as a set of a single dot and circles of different size, constructing the backbone of the manifold.
The disturbed dynamics shown as gray trajectories reside around the backbone, which describes the effect of vortex gust about the baseline dynamics in a low-rank manner.
This physically-interpretable shape suggests that the CNN-AE learns the important vortical structures associated with lift through the observable augmentation, which is inspired by the aerodynamic prior knowledge of $\Gamma \propto C_L$ where $\Gamma$ is a circulation.
We note that the discovered cone shape can call for phase-amplitude modeling, enabling the optimal control design in the latent space, which will be discussed in section~\ref{sec:FC}.

\begin{figure}[b!]
    \centering
    \includegraphics[width=0.85\textwidth]{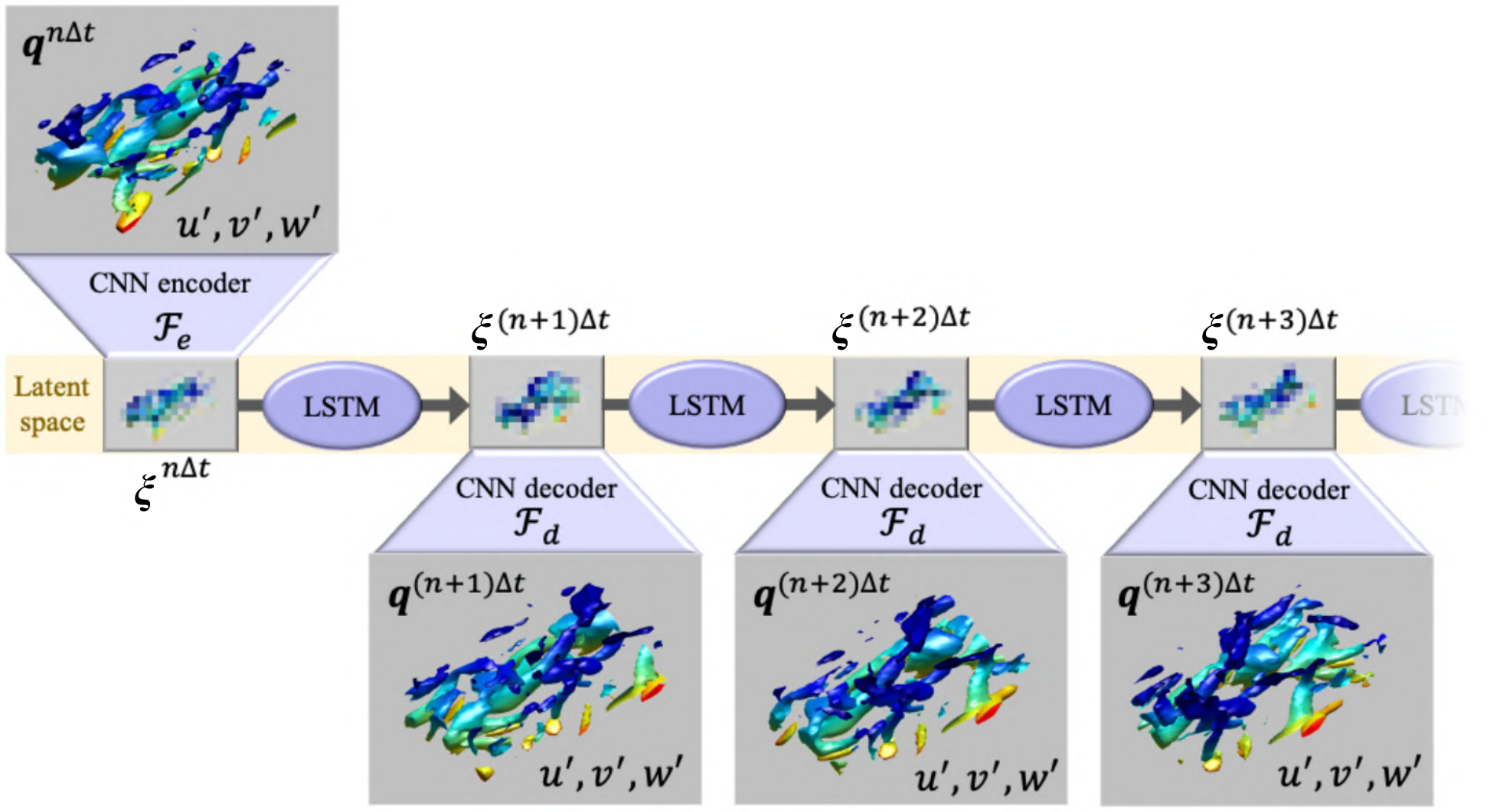}
    \caption{
    A general framework of latent reduced-order modeling.
    Once a flow field snapshot is compressed into the latent vector, the temporal evolution of latent vector is predicted through different machine-learning techniques.
    An LSTM is used as a latent temporal integrator in this example.
    (Adapted from:
    Nakamura T, Fukami K, Hasegawa K, Nabae Y and Fukagata K 2021 
    {\it Phys.~Fluids} {\bf 33} 025116.
    Copyright \copyright~2021 Authors.)    
    }
    \label{fig_Taichiman}
\end{figure}

The framework of observable-augmented autoencoder can be generalized by replacing the input/output of the autoencoder and a given observable depending on the problem of interest.
There are indeed a number of follow-up studies using the observable-augmented CNN-AE for, e.g., sparse sensor-based reconstruction~(Fukami and Taira 2024a, Mousavi and Eldredge 2025, Eldredge and Mousavi 2025), data fusion~(Fukami and Taira 2025), shape optimization~ (Tran {\it et al} 2024), and reinforcement learning-based control (Liu {\it et al} 2024).
Rather than na\"ively performing CNN-AE compression, seeking such a physically-interpretable latent representation could offer new perspectives of nonlinear autoencoder-based mode decomposition for unsteady flows.

\subsection{Latent modeling}
\label{sec:LM}


Challenges we always face with fluid flow phenomena come from their high dimensional and nonlinear natures.
Even if we can perform large-scale simulations or high-definition measurements, this would remain in interpreting the obtained data with million, billion, or trillion degrees of freedom.
Therefore, reduced-order models, which can represent important features of flows and their dynamics with a much lower degree of freedom, are useful for understanding complex flow phenomena and designing effective control laws.

Whether linear or nonlinear, a typical reduced-order model is composed of a spatial compressor, which is expressed in the same as equations (1) and (2), and a temporal integrator predicting the dynamics in the low-dimensionalized latent space, i.e., 
\begin{equation}
\frac{{\rm d}\bm{\xi}}{{\rm d}t} = {\cal A}(\bm{\xi}) ,
\end{equation}
where $\cal A$ is the operator describing the latent dynamics.
Within the linear methods, POD is widely used as a spatial compressor, i.e.,
\begin{equation}
\bm{q}(\bm{x},t) = \sum_{j=0}^{n_{\bm{\xi}}} \xi_j (t) \bm{\Psi}_j (\bm{x}), 
\end{equation}
where $\bm{\Psi}_j$ denotes the $j$-th POD mode.
The Galerkin projection can be then used to construct the low-dimensional dynamical system.
For instance, for incompressible flows, $i$-th component of $\cal A$ is expressed as
\begin{equation}
{\cal A}_i = 
\sum_{j=0}^{n_{\bm{\xi}}} \sum_{k=0}^{n_{\bm{\xi}}} A_{ijk} \xi_j \xi_k
+
\sum_{j=0}^{n_{\bm{\xi}}} D_{ij} \xi_j,
\end{equation}
where $A_{ijk}=-\int_V \bm{\Psi}_i \cdot (\bm{\Psi}_j \cdot \nabla \bm{\Psi}_k) {\rm d}V$ and 
$D_{ij} = {\rm Re}^{-1}\int_V \bm{\Psi}_i \cdot \nabla^2 \bm{\Psi}_j {\rm d}V$ are the Galerkin projections of the advection and diffusion terms, respectively.

A similar routine can be considered in constructing reduced-order models with machine learning (hereafter referred to as ML-ROM).
Machine learning can be used as a spatial compressor, a temporal integrator, or both, as illustrated in figure~\ref{fig_Taichiman}.
Once a flow field snapshot is compressed into the latent vector, the temporal evolution of the latent vector is predicted using a temporal integrator.
For example, Hasegawa {\it et al} (2020a, 2020b) used a CNN-AE as a spatial compressor and a long-term short memory (LSTM; Hochreiter and Schmidhuber 1997) as a temporal integrator to construct an ML-ROM; namely,
\begin{equation}
\bm{\xi}^{n+1} = {\cal F}_{\rm LSTM}(\bm{\xi}^{n}, \bm{\xi}^{n-1}, \cdots) .
\end{equation}
They examined its capability for unsteady flows around bluff bodies of different shapes and flows around a circular cylinder at different Reynolds numbers.

\begin{figure}[b!]
    \centering
    \includegraphics[width=\textwidth]{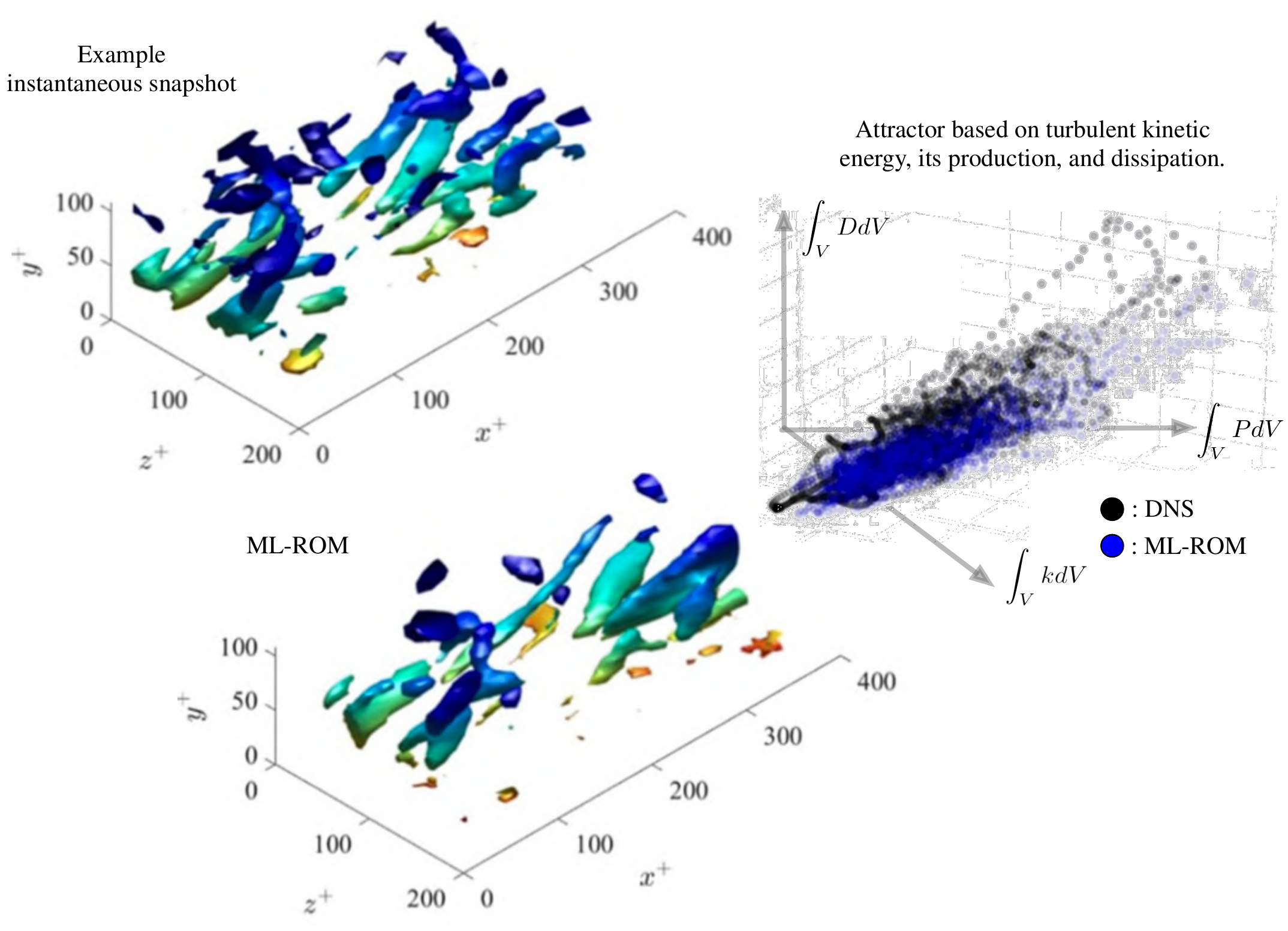}
    \caption{
    Comparison of flow field and attractor between the reference simulation and machine-learning-based latent reduced-order surrogate modeling of turbulent channel flow. 
    (Adapted from:
    Nakamura T, Fukami K, Hasegawa K, Nabae Y and Fukagata K 2021 
    {\it Phys.~Fluids} {\bf 33} 025116.
    Copyright \copyright~2021 Authors.) 
    }
    \label{fig_Taichi2}
\end{figure}

Nakamura {\it et al} (2020, 2021) extended this CNN-AE/LSTM-based ML-ROM to deal with a three-dimensional velocity field in a turbulent channel flow.
While the case of laminar vortex shedding can be accurately predicted with a standard formulation of CNN and LSTM (Hasegawa {\it et al} 2020a, 2020b), they found that both ML models need to be carefully designed to learn the intrinsic features of wall-bounded turbulence.
The CNN-AE model includes multiple filters with different sizes referred to as a multi-scale CNN-AE~(Fukami {\it et al} 2019a, 2020a) to extract a range of spatial length scales residing in a channel flow.
To capture the different temporal scales of the turbulent flow depending on its dynamical motion, multiple LSTMs with different numbers of units are used in parallel.
More details of this turbulence-oriented model construction are referred to Nakamura {\it et al} (2020).

Figure~\ref{fig_Taichi2} shows the vortical structure in a turbulent flow in a minimal channel at ${\rm Re}_\tau =110$ (Nakamura {\it et al} 2020, 2021), which demonstrates that the vortical structure of DNS is reasonably well reproduced by ML-ROM.
The trajectories in a three-dimensional phase space of the production $P$, the turbulent kinetic energy $k$, and the dissipation $D$ are also presented in figure~\ref{fig_Taichi2}.
The region where the trajectory reproduced by the ML-ROM resides largely overlaps with that of the DNS, indicating that the attractor embedded in this flow is well captured by the proposed ML-ROM. 
However, a strongly intermittent event far from the basin of the attractor is not sufficiently captured, implying a need of further improvement of the model.

In contrast to the POD Galerkin model, a machine-learning-based temporal integrator becomes a black box when constructed with neural networks.
Toward facilitating the interpretation of what machine-learning models learn over dynamics and extending the model to flow control, one can consider using a gray-box model such as sparse identification of nonlinear dynamics (SINDy; Brunton {\it et al} 2016).
This provides an explicit expression of modeled equations in the latent space.
For example, Fukami {\it et al} (2021c) used SINDy for the low-dimensionalized field obtained by a CNN-AE.
They examined SINDy with different regression methods and parameter choices for a two-dimensional single cylinder wake at ${\rm Re}_D = 100$, its transient process, and a wake of two-parallel cylinders.

\begin{figure}[b!]
    \centering
    \includegraphics[width=\textwidth]{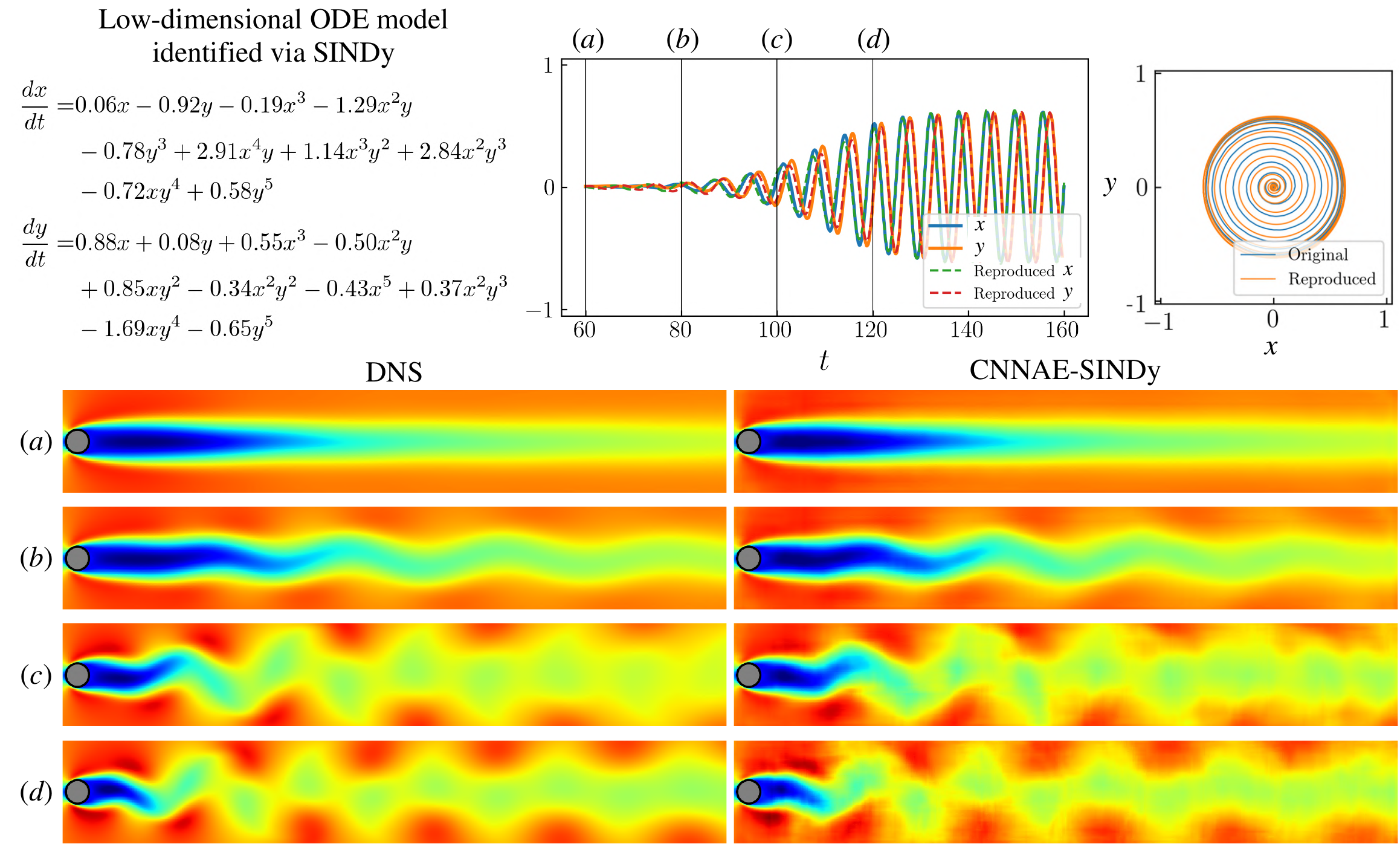}
    \caption{
    SINDy-based latent dynamical modeling of transient process of flow around a cylinder.
    The identified model equation, the time trace of latent vector, and the temporal evolution of the reproduced wake field are depicted.
    (Adapted from:
    Fukami K, Murata T, Zhang K and Fukagata K 2021 {\it J. Fluid Mech.}~{\bf 926} A10.
    Copyright \copyright~2021 Author(s), CC BY 4.0 license.) 
    }
    \label{fig_AESINDy}
\end{figure}

An example set of identified modeled equations for the two-dimensional latent variables $\bm{\xi} = (x,y)$ is presented in figure~\ref{fig_AESINDy}~(Fukami {\it et al} 2021c).
The figure also shows the time traces and the trajectory of $(x,y)$ reproduced by temporally integrating this set of equations, and the transient velocity fields obtained by decoding those $(x,y)$.
These variables are in agreement with the original ones from DNS, thereby successfully reproducing the transient dynamics of the cylinder wake from two latent variables only.
Although not shown here, they also studied a nine-equation turbulent shear flow model to examine the applicability of SINDy to turbulence. 
The result suggested that a similar reduced-order modeling is possible once the low-dimensionalization is successfully performed.
While this latent modeling using SINDy has been shown to work well, they have cautioned that choices of the regression method and the selection parameter are very important for identifying a reasonably sparse yet accurate expression of the latent dynamics.
Otherwise, the obtained equations may be too complicated or inaccurate.

As a fluid mechanician, the applicability of such ML-ROMs to numerical simulation may be of interest.
There are a good number of attempts in this aspect; however, it is still challenging to combine with or support high-fidelity numerical simulation (Vinuesa and Brunton 2022).
One reason is that errors accumulated through ML-ROM operations such as CNN-AE-based compression and temporal prediction in the latent or physical space lead to an unstable calculation or inaccurate flow prediction with respect to the conservation laws and boundary conditions.

\begin{figure}[b!]
    \centering
    \includegraphics[width=\textwidth]{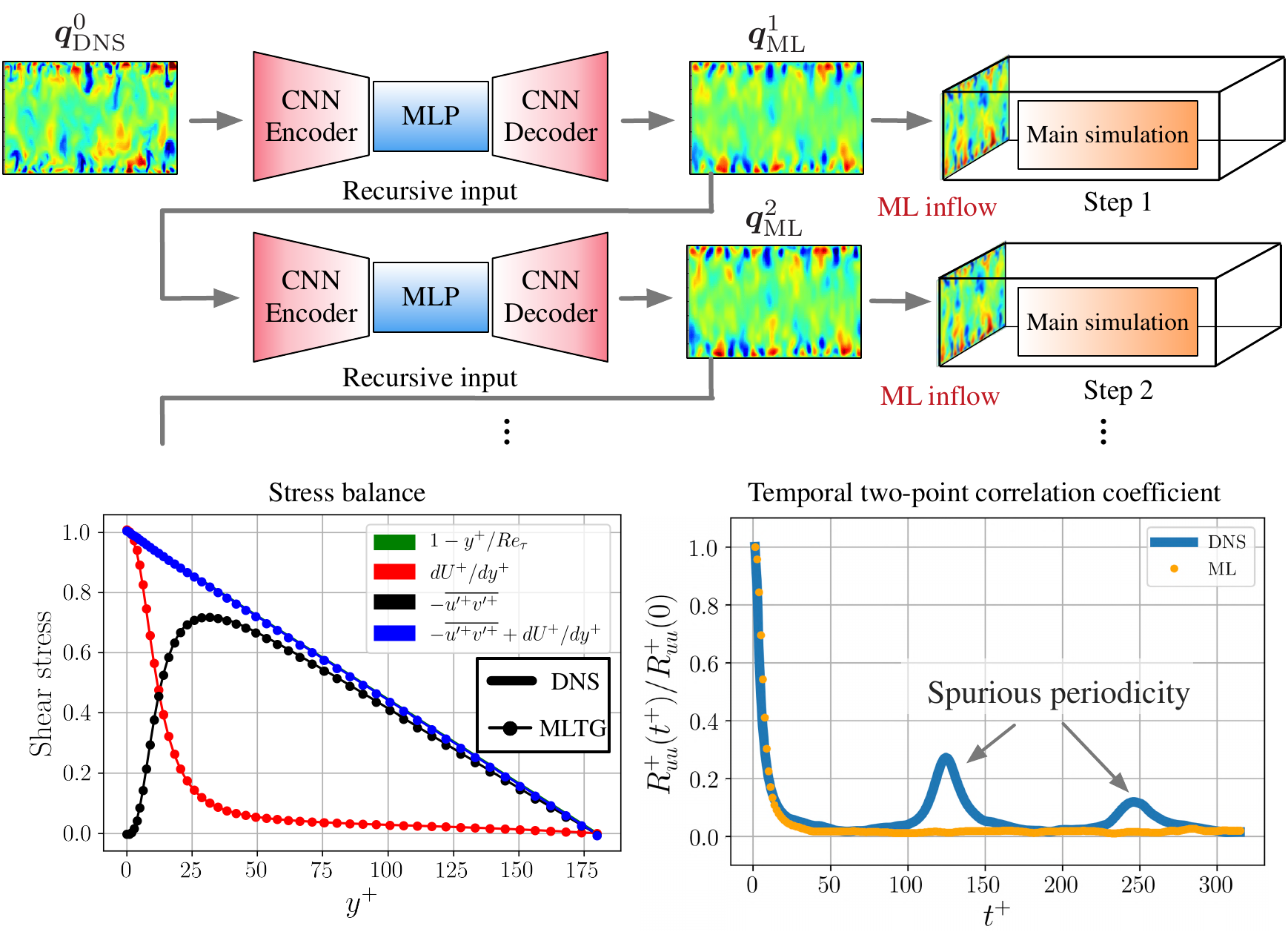}
    \caption{
    Latent modeling-based inflow condition generator for a numerical simulation of turbulent channel flow. 
    Statistics measured from the machine-learning-assisted simulation ({\it a posteriori} test) are also shown.
    (Adapted from: Fukami K, Nabae Y, Kawai K and Fukagata K 2019 {\it Phys.~Rev.~Fluids} {\bf 4} 064603.
Copyright \copyright~2019 American Physical Society. APS copyright policies allow the authors to reuse figures without written permission.)
    }
    \label{fig_MLTG}
\end{figure}

Keep the above point in mind, let us introduce one of the pioneering studies for machine-learning-based turbulent flow prediction aiming to support direct numerical simulation by Fukami {\it et al} (2019b).
A direct regression between the velocity fields at time steps $n$ and $n+1$ is performed by using a network whose structure is the same as that of CNN-AE.
They considered replacing conventional driver simulations or synthetic turbulent inflow generators with this methodology of temporal predictor for turbulent flow, referred to as machine-learned turbulent inflow generator (MLTG), as illustrated in figure \ref{fig_MLTG}.

The CNN-AE-type model is trained using instantaneous velocity fields in a single cross-section obtained DNS of a fully developed turbulent channel flow at the friction Reynolds number of ${\rm Re}_\tau = 180$~(Fukami {\it et al} 2019b).
The cross-sectional velocity field at time $n+1$ is predicted from that at time $n$, 
\begin{equation}
\bm{q}^{n+1} = {\cal F}_{\rm ML}(\bm{q}^{n}) ,
\label{eq:MLTG}
\end{equation}
where $\bm{q}$ is composed of the three components of velocity fluctuations.
Note that, unlike the ML-ROMs introduced above, the spatial compressor and the temporal integrator are mixed in the single model ${\cal F}_{\rm ML}$.
Once the model is trained, the temporal evolution of the velocity field is computed by recursively applying equation (\ref{eq:MLTG}), although the initial condition needs to be prepared by using, e.g., DNS data $\bm{q}^0_{\rm DNS}$.
Example movies of the predicted flow fields over time with the recursive temporal prediction are available in \url{https://kflab.jp/en/index.php?MLTG2}.

The shear stress profiles obtained by this MLTG are in agreement with the reference DNS, as presented in figure \ref{fig_MLTG}. 
The other turbulence statistics obtained through the inflow-outflow DNS with this MLTG at the inlet have also been confirmed to be in excellent agreement with those of conventional inflow-outflow DNS with a driver (or precursor) DNS, while the computational time was reduced to about 1/580 under the authors' environment at that time (NVIDIA Tesla K40 GPU).
In addition, this MLTG is free from spurious periodicity, which is a problematic issue when the conventional driver DNS in a periodic domain is used (Wu 2017), as shown in figure~\ref{fig_MLTG}.

\begin{figure}[b]
    \centering
    \includegraphics[width=0.7\textwidth]{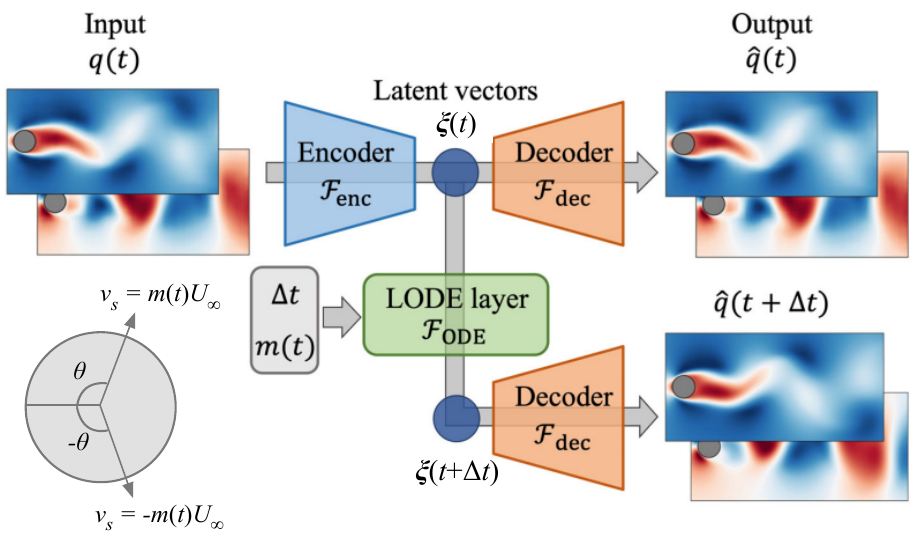}
    \caption{
    Linear extraction system autoencoder.
    (Adapted from: Ishize T, Omichi H and Fukagata K 2024 {\it Int.~J.~Numer.~Methods Heat Fluid Flow} {\bf 34} 3253--3277.
Copyright \copyright~2024 Emerald Publishing Limited. 
Emerald's copyright policies allow the authors to reuse figures without written permission.)
    }
    \label{fig_LEAE}
\end{figure}

We again note that careful use of ML-ROM is required to accordingly support high-fidelity simulation while ensuring the validity of outcomes from the numerical simulation.
The study by Fukami {\it et al} (2019b) above also reported that a post-processing for the predicted velocity field is needed to stabilize the main simulation in accelerating it with the MLTG. 
A similar observation is also seen in machine-learning-based turbulence modeling (Park and Choi 2021).
A remedy may be the use of physics-informed learning that accounts for governing equations and high-order statistics in the cost function of machine-learning model training (Raissi {\it et al} 2019)

From the aspect of surrogate modeling, the usability of ML-ROMs may also depend upon the flow of interest.
For example, while long-term dynamics may become interested in statistically-steady flows, an accurate instantaneous prediction would be necessary for transient flows or systems with extreme events.
Hence, a temporal predictor of ML-ROMs can be chosen based on the characteristics of machine-learning models and a target flow users would consider. 
With the recent developments of machine-learning-based flow prediction models such as transformer (Yousif {\it et al} 2023, Solera-Rico {\it et al} 2024), echo state network (Racca {\it et al} 2023), and neural ODE
 (Linot {\it et al} 2023a), CNN-AE-based latent modeling may support a range of unsteady flow studies.

\subsection{Flow control}
\label{sec:FC}

In addition to modeling and prediction, controlling fluid flows is of great engineering importance.
This enables us to address a range of engineering problems, including drag reduction, mixing enhancement, and noise alleviation, all of which are necessary elements for a sustainable society (Kasagi {\it et al} 2009).
Among different flow control strategies, 
feedback control has been extensively studied
(Abergel and Temam 1990, Kim and Bewley 2007).
However, applying such control theories to fluid flow problems is often challenging or, even if possible, computationally costly due to their high-dimensional and nonlinear nature.
Therefore, it may be advantageous to easily apply control theory to fluid flows if the flow systems are described by a low-dimensional model through an ML-ROM.

\begin{figure}[b!]
    \centering
    \includegraphics[width=\textwidth]{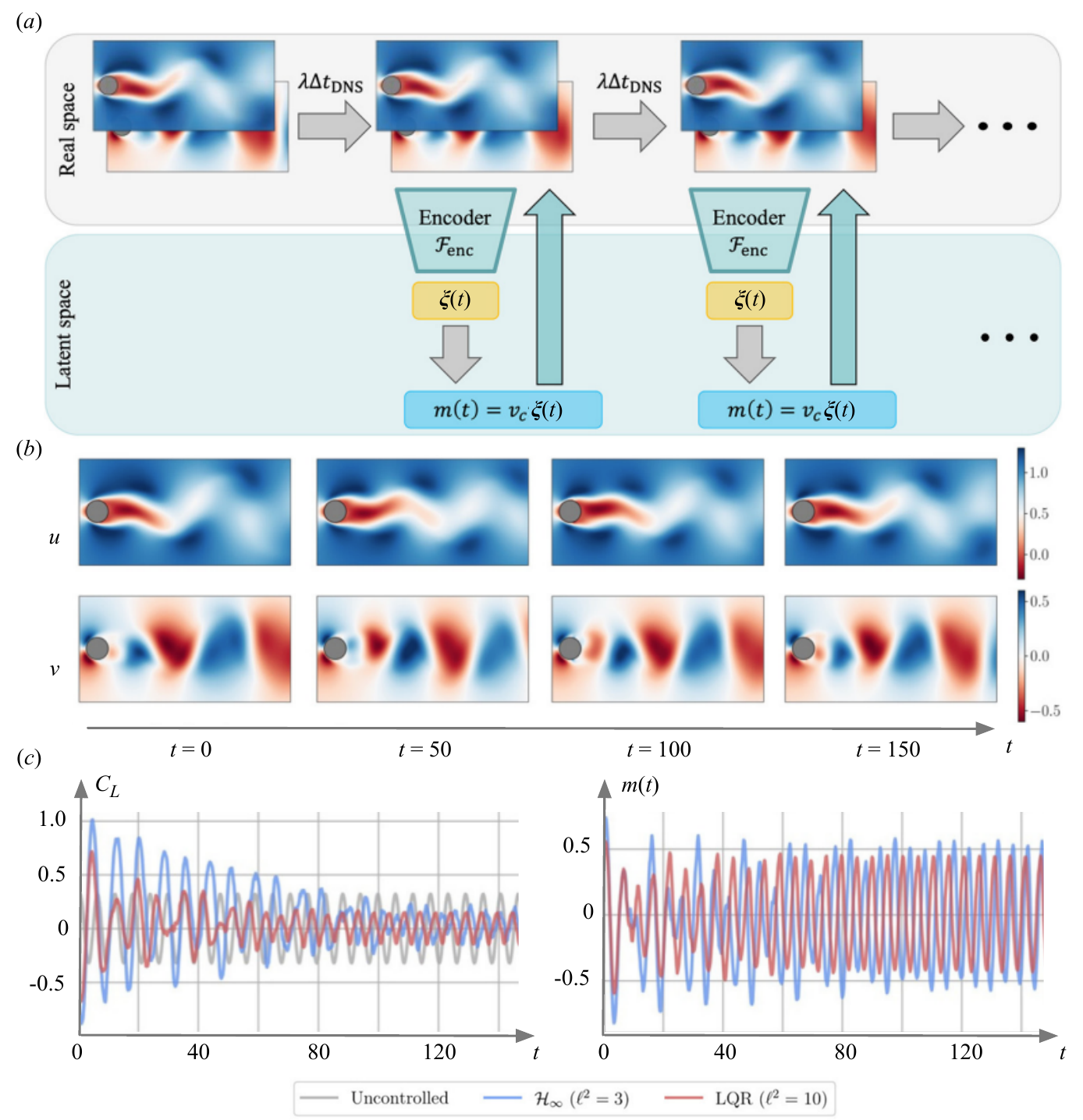}
    \caption{
    Linear extraction system autoencoder-aided control design for shedding suppression of laminar cylinder wake.
    $(a)$ Applying the control input designed in the low-order subspace to the flow field.
    Time series of $(b)$ the controlled flow field, $(c)$ lift coefficient and designed control input.
    (Adapted from: Ishize T, Omichi H and Fukagata K 2024 {\it Int. J. Numer. Methods Heat Fluid Flow} {\bf 34} 3253–-3277.
Copyright \copyright~2024 Emerald Publishing Limited. 
Emerald's Copyright Policies allow the authors to reuse figures without written permission.)
    }
    \label{fig_LEAE_res}
\end{figure}

One way to build a bridge between control theories and fluid flow systems is to extract linearized, low-order dynamics through an autoencoder
 (Ishize {\it et al} 2024, Lusch {\it et al} 2018).
Here, let us introduce such an attempt by Ishize {\it et al} (2024), proposing the linear extraction system autoencoder (LEAE) as illustrated in figure~\ref{fig_LEAE}.
They considered a two-dimensional flow around a circular cylinder at ${\rm Re}_D=100$ controlled using the blowing and suction slots located at $\pm 110^\circ$ from the front stagnation point similar to previous studies
(Park {\it et al} 1994, Illingworth {\it et al} 2014).

The control target is to suppress vortex shedding by a feedback control of blowing/suction amplitude $m(t)$.
The LEAE consists of CNN-AE and a custom layer referred to as a linear ordinary differential equation (LODE) layer.
This is designed to integrate the state equation in the latent space,
\begin{equation}
\frac{{\rm d}\bm{\xi}}{{\rm d}t} = \bm{A}\bm{\xi} + \bm{B}\bm{m} ,
    \label{eq:LE}
\end{equation}
with the Crank-Nicolson scheme; {\it viz.},
\begin{equation}
    {\bm \xi}(t+\Delta t)=(2{\bm I}-\Delta t{\bm A})^{-1}
    [(2{\bm I}+\Delta t{\bm A}){\bm \xi}(t)+\Delta t {\bm B} ({\bm m}(t)+{\bm m}(t+\Delta t))],
    \label{eq:CN_ODE2}
\end{equation}
where $\Delta t$ is the time step size.
The system matrix $\bm{A}$ and the control matrix $\bm{B}$ are identified through the training using the training data consisting of various cases of blowing/suction amplitudes.
The weights inside the decoders for the $n$ time step and the $n+1$ time step are shared.
The unique feature of this model is that the weights in the CNN-AE and the matrices $\bm{A}$ and  $\bm{B}$ in the LODE layer are trained simultaneously, while the structure of the LODE layer is fixed as in equation (\ref{eq:CN_ODE2}).
In other words, this CNN-AE is trained under the constraint such that the latent dynamics should follow the linear equation (\ref{eq:LE}).

Once the system and control matrices $\bm{A}$ and $\bm{B}$ are identified such as
\begin{equation}
\frac{{\rm d}\bm{\xi}}{{\rm d}t} =
\underbrace{
\left[
\begin{array}{cc}
0 &-1.066\\
1.066& 0\\
\end{array}
\right]
}_{\bm{A}}
{\bm \xi}(t)+
\underbrace{
\left[
\begin{array}{cc}
{-0.03672}\\
{0.02672}\\
\end{array}
\right]
}_{\bm{B}}
{m}(t),
\label{eq:Linear_system_wcont}
\end{equation}
derivation of the control law is straightforward using the control theory.
In their study, the linear quadratic regulator (LQR) and the ${\cal H}_{\infty}$ control are considered as representative control methods. 
The control law is expressed as
\begin{equation}
{\bm m}(t) = {\bm {K}}\bm {\xi}(t),
\label{eq:min_J}
\end{equation}
and the feedback gain $\bm{K}$ is computed by
\begin{equation}
{\bm K} = -{\frac{1}{\ell^2}}{\bm {BP}},
\label{eq:gain}
\end{equation}
where ${\ell^2}$ is the parameter for the price of control, and $\bm{P}$ is the solution of the corresponding Riccati equation.
More details are referred to Ishize {\it et al} (2024).

\begin{figure}[b!]
    \centering
    \includegraphics[width=0.65\textwidth]{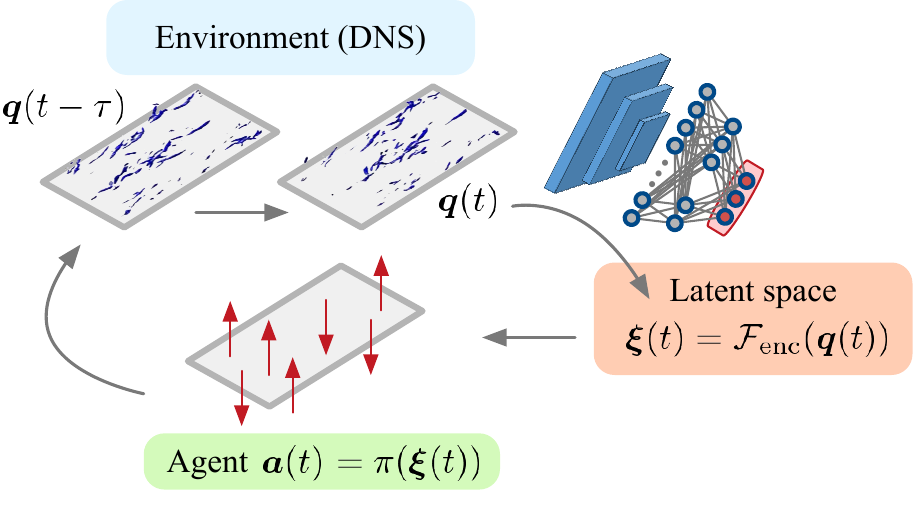}
    \caption{
    Reinforcement learning for control design in the latent space.
    }
    \label{fig_RLlat}
\end{figure}

The control results with the LQR and ${\cal H}_{\infty}$ control laws derived using LEAE are summarized in figure~\ref{fig_LEAE_res}.
In this example, the flow in the real space is simulated by DNS.
As shown in figure~\ref{fig_LEAE_res}($a$), the flow is compressed using the encoder part of the LEAE at a certain time interval $\lambda\Delta t_{\rm DNS}$ ($\lambda=100$ in this case) to obtain the latent vector $\bm{\xi}(t)$.
The control input computed by equations (\ref{eq:min_J}) and (\ref{eq:gain}) is then fed back to the real space.
Namely, the procedure for determining the control input is summarized as
\begin{equation}
m(t) = {\bm{K}}
{\cal F}_{\rm enc}({\bm q}(t)).
\label{eq:cont_law_DNS}
\end{equation}

The time variations of the streamwise ($u$) and transverse ($v$) velocity components from the time instant $t=0$ when the control is turned on are presented in figure~\ref{fig_LEAE_res}($b$).
Although the control effect is not that obvious from these visualizations, the time trace of the lift coefficient $C_L$ computed from these velocity fields shown in figure~\ref{fig_LEAE_res}($c$) exhibits the suppression of vortex shedding.

In addition to involving conventional control theories in the latent space, one can consider integrating a modern machine-learning technique of reinforcement learning in a low-order space to design the optimal control law in a feedback manner.
The objective of reinforcement learning is to optimally derive the time-varying control policy (or action) $a(t) = \pi({\bm q}(t))$ for a given state ${\bm q}$.
However, as mentioned in section \ref{sec:AEg}, this state ${\bm q}$ in fluid mechanics problems generally corresponds to a flow field possessing a large degree of freedom.
Because reinforcement learning requires collecting the time-series information of states in training an agent $a(t)$, the aforementioned largeness of state variable ${\bm q}$ is a unique bottleneck when applying reinforcement learning for flow control.
In response, Linot {\it et al} (2023b) proposed an autoencoder-assisted reinforcement learning control of unsteady flows, which uses latent variables ${\bm \xi}(t)$ as a state of reinforcement learning routine instead of the original flow field ${\bm q}(t)$, as illustrated in figure~\ref{fig_RLlat}.

\begin{figure}[b!]
    \centering
    \includegraphics[width=0.85\textwidth]{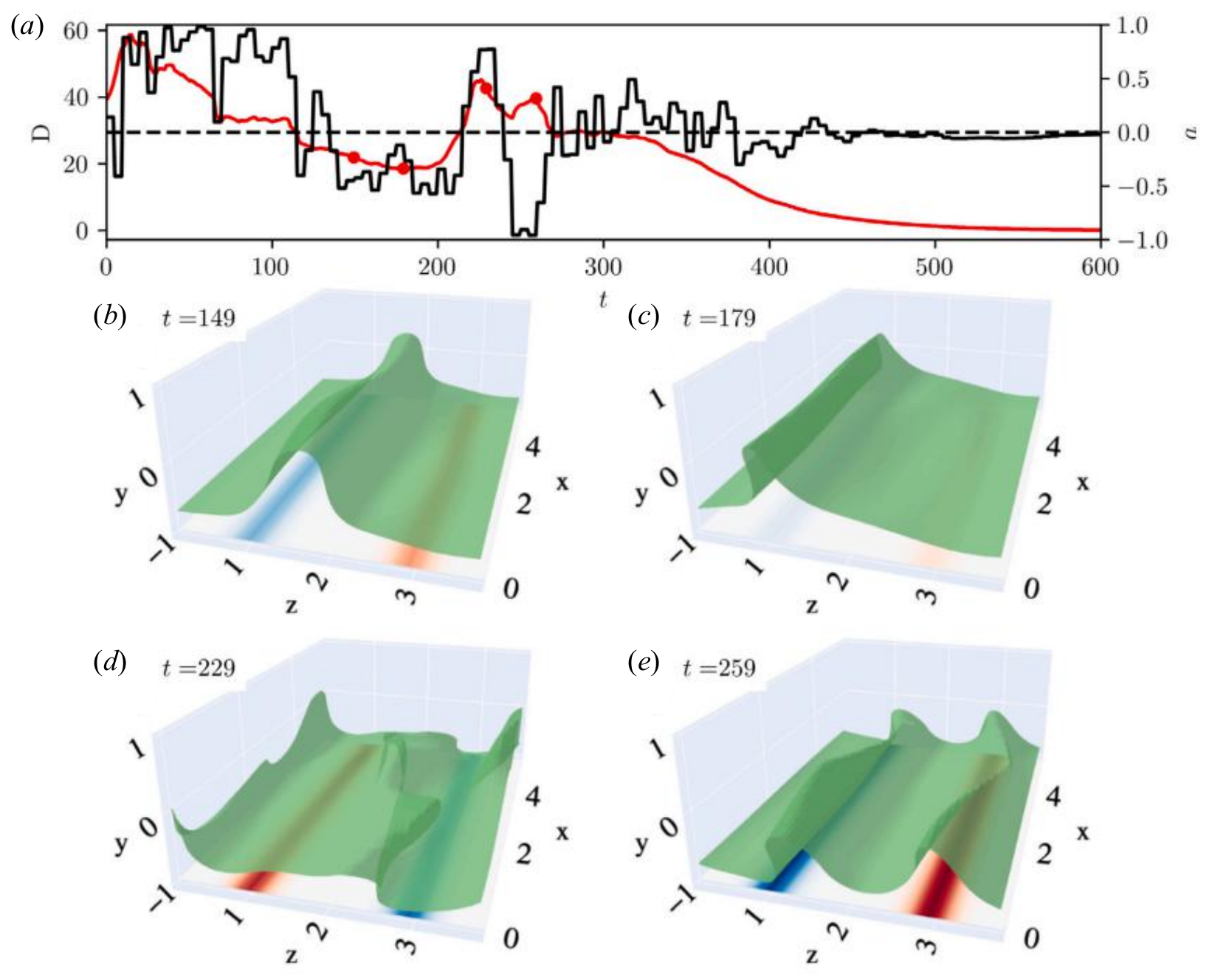}
    \caption{
    Reinforcement learning-based latent control design for plane Couette flow.
    $(a)$ Time trace of drag $D$ (colored in red) and action $a$ (colored in black).
    $(b)-(e)$ Representative flow snapshots over time.
        (Adapted from: Linot A J, Zeng K and Graham M D 2023 
        {\it Int.~J.~Heat Fluid Flow} {\bf 101} 109139.
Copyright \copyright~2023 Elsevier Inc. With permission.)
    }
    \label{fig_Alec}
\end{figure}

Considering the Couette flow with two slot jets on one wall, they performed the reinforcement learning-based latent control design in the latent space obtained from a nonlinear autoencoder, as presented in figure~\ref{fig_Alec} (Linot {\it et al} 2023b).
Here, the domain size and the Reynolds number of 400 for the analysis are carefully chosen to sustain a self-sustaining process driving wall-bounded turbulence. 
The control objective is to minimize the turbulent drag $D$ averaged between both the top and bottom walls
subject to a quadratic penalty on actuation amplitude $a(t)$.
The actuation is performed on the bottom wall, possessing the form of two slot jets with the Gaussian profile in the spanwise direction, denoted as $f_a(x,z) = a(t)f(x,z)$.
Hence, the reinforcement learning agent is trained to optimize the action $a(t)$ to minimize the drag $D$ based on the low-order latent representation ${\bm \xi}(t)$ with 25 dimensions that are much less compared to the original data dimension of the flow state ${\bm q}$.

The turbulent drag $D$ colored in red in figure~\ref{fig_Alec}$(a)$ is successfully reduced over time.
They also found that the agent trained in the latent space provides a novel drag-reduction strategy of manipulating the low-speed streak to a preferred location, leading to the breakdown of the streak, and in the wake of the break-down forming two low-speed streaks in its place
(Linot {\it et al} 2023b).
As these two low-speed streaks are unsustainable within the domain, this results in breaking the self-sustaining process thereby achieving the laminarization, as presented in figures~\ref{fig_Alec}$(b)-(e)$.
Reinforcement learning in the latent space may provide an effective, nonlinear control policy with low cost, which is often challenging to find by only relying on human insights.
We could also learn flow physics based on the time-varying optimized action derived through reinforcement learning.

\begin{figure}[b!]
    \centering
    \includegraphics[width=\textwidth]{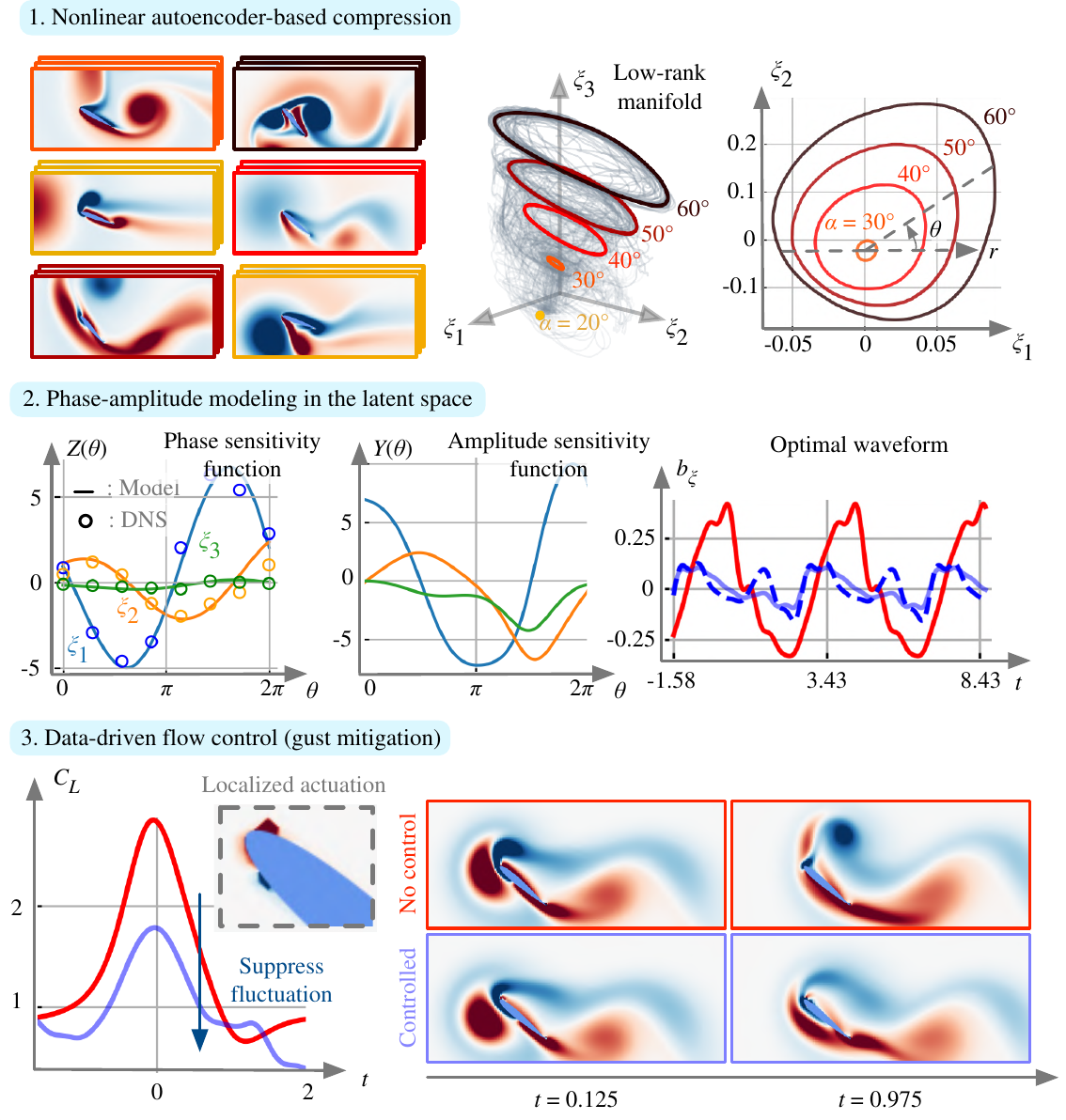}
    \caption{
    Phase-amplitude modeling-based latent control design for extreme vortex-airfoil interactions.
    (Adapted from:
    Fukami K, Nakao H and Taira K 2024 {\it J. Fluid Mech.}~{\bf 992} A17.
    Copyright \copyright~2024 Author(s), CC BY 4.0 license.) 
    }
    \label{fig_Phase}
\end{figure}

While the above reinforcement learning strategy is powerful in finding the optimal feedback control law, it still needs a large data set to make the latent-space model robust against various actuated scenarios in addition to the baseline dynamics.
Let us lastly introduce an example of feedforward flow control in the latent space (Fukami {\it et al} 2024b), which may enable reducing the burden of machine-learning-based flow control.

We here revisit the cone-shape submanifold of extreme vortex-airfoil interaction identified through a lift-augmented autoencoder discussed in section~\ref{sec:MD}.
As the backbone of the manifold is composed of a set of circles, the latent dynamics can also be described through the phase $\theta$ and amplitude $r$, as presented in figure~\ref{fig_Phase}.
For such phase-amplitude dynamics, phase-amplitude reduction analysis can be considered to examine the system response while deriving the control strategy of given dynamics (Nakao 2021).

Concretely, the system response is provided as a sensitivity function of phase and amplitude.
The phase sensitivity function ${\bm Z}(\theta)$ describes the sensitivity of the phase of the dynamical system and the amplitude sensitivity function ${\bm Y}(\theta)$ expresses the sensitivity of the system amplitude about the periodic orbit against external actuation inputs.
Although there are a couple of ways to measure the sensitivities ${\bm Z}(\theta)$ and ${\bm Y}(\theta)$, one can obtain them by assessing the left Floquet eigenvectors for a dynamical model (Takata {\it et al} 2021).
Once the sensitivities are available, the optimal actuation pattern can be derived through optimization with respect to phase and amplitude.

As the fluctuation of aerodynamic loads due to the vortex impingement happens within a short time such as one convective time unit, it is preferable to swiftly mitigate the impact of gusts. 
On the identified cone-shape submanifold, the phase and amplitude respectively describe the timing and effect of vortex gusts about the undisturbed baseline dynamics (Fukami {\it et al} 2024b).
Inspired by this geometric shape of the latent submanifold reflecting the key feature of high-dimensional vortical flows, the control objective is set to reduce the amplitude deviation corresponding to the gust effect while quickly modifying the phase.
Once the latent dynamical model $\dot{\bm \xi}$ is identified using SINDy (Brunton {\it et al} 2016), the phase and amplitude sensitivity functions are analytically obtained, thereby providing the optimal-synchronization waveform with amplitude suppression in the low-order latent space, as depicted as the second step in figure~\ref{fig_Phase}.

The data-driven lift attenuation control for an example case of strong positive vortex-wing interaction is presented at the bottom portion in figure~\ref{fig_Phase}.
Here, they employ the body force near the wing, whose shape is obtained through the Jacobian of the CNN encoder.
Details are referred to Fukami {\it et al} (2024b).
The vortex gust is shifted downward and then slips over the pressure side of the wing due to the actuation, resulting in significant suppression of the lift fluctuation.
Furthermore, this is achieved within sole one convective time, suggesting that the low-order latent space obtained through the lift-augmented CNN-AE captures the essence of high-dimensional vortical flows.

A follow-up study with particle image velocimetry data has further shown that the determination of phase for unsteady flows in the latent space is possible even for non-periodic flows with a variant of CNN-AE, referred to as a topological autoencoder (Smith {\it et al} 2024).
Equipped with these approaches, nonlinear CNN-AE compression and the resulting latent representation ${\bm \xi}$ may offer insights into understanding and controlling a range of unsteady flows, which goes beyond what has been possible by na\"ively gazing at a high-dimensional flow state~$\bm{q}$.

\section{Summary and Outlook}

{
In the present review, we have introduced the fundamentals of flow field compression using autoencoders --- in particular, convolutional neural network-based autoencoders (CNN-AE) --- and their applications to various fluid dynamics problems.
While the minimization problem for autoencoder is similar to that of the proper orthogonal decomposition (POD), the nonlinear activation function in the autoencoder improves the compression ability, which is especially beneficial for handling fluid flow data.

From the applications of CNN-AE to different flow problems, we have seen that CNN-AE can successfully be used for mode decomposition and latent modeling, leading to the construction of machine learning-based reduced-order models (ML-ROMs).
Such an ML-ROM can be used as a surrogate model for reducing the computational cost of some simulations such as the turbulent-inflow generator.
It is also possible to design control strategies of unsteady flows based on the knowledge in the latent space.
}

{
As discussed throughout the paper, what an autoencoder extracts as a low-order representation in the latent subspace depends on how we set the optimization problem or cost function.
Adding some constraints based on prior knowledge of control theory and physics, such as the constraint to follow a linear system (Ishize {\it et al} 2024) and that to produce the aerodynamic coefficient (Fukami and Taira 2023) introduced here has shown new possibilities of autoencoder-based analysis. 
Other types of prior knowledge-based design for cost function or model construction of autoencoder may yield new insights into high-dimensional unsteady flow physics in a compact manner.

Furthermore, how we smartly learn flow physics from data with respect to data amount and fidelity throughout CNN-AE would be of interest in the future.
While a range of numerical and experimental flow data become made available nowadays (Towne {\it et al} 2023), extracting flow characteristics from various sources and fidelity of data needs to be considered.
As experiments can typically achieve higher Reynolds numbers compared to numerical simulations, this type of data fusion could enable feature extractions from a wide range of Reynolds numbers (Fukami and Taira 2025).
In addition, since we know there exists some level of spatiotemporal universality such as scale invariance of fluid flows (Fukami and Taira 2024b), a collection way of training data could also be revisited rather than na\"ively sampling available big data.
These arguments hold also when we want to reconstruct flow fields from a very limited number of sensors in natural environments (Oura {\it et al} 2024).
Autoencoder constructions with such flow physics-oriented thinking may be needed for supporting unsteady flow analysis in a data-driven manner.
}

\section*{Acknowledgments}
This review is an extension of the invited talk given at ICTAM2024 (Fukagata 2024).
This work was generously supported by JSPS KAKENHI Grant Numbers JP18H03758 and JP21H05007.
We thank our collaborators for CNN-AE-related studies of unsteady fluid flow problems, including
Kazuto Hasegawa,
Takeru Ishize,
Shoei Kanehira,
Ken Kawai, 
Soshi Kawai, 
Sangseung Lee,
Romit Maulik,
Masaki Morimoto,
Takaaki Murata,
Yusuke Nabae,
Aditya Nair,
Taichi Nakamura,
Hiroya Nakao,
Hiroshi Omichi,
Luke Smith,
Kunihiko Taira,
Jonathan Tran,
Ricardo Vinuesa,
Koichiro Yawata,
and 
Kai Zhang,
for fruitful discussions.

\section*{ORCID iDs}
Koji Fukagata~  \url{https://orcid.org/0000-0003-4805-238X}\\
Kai Fukami~  \url{https://orcid.org/0000-0002-1381-7322}



\end{document}